Corresponding Author: Mr. Inaki Ordonez-Etxeberria,

Corresponding Author's Institution: University of the Basque Country

First Author: Inaki Ordonez-Etxeberria

Order of Authors: Inaki Ordonez-Etxeberria; Ricardo Hueso; Agustin Sánchez-Lavega; Santiago Pérez-Hoyos

Abstract: The Cassini spacecraft made a gravity assist flyby of Jupiter in December 2000. The Imaging Science Subsystem (ISS) acquired images of the planet that covered the visual range with filters sensitive to the distribution of clouds and hazes, their altitudes and color. We use a selection of these images to build high-resolution cylindrical maps of the planet in 9 wavelengths. We explore the spatial distribution of the planet reflectivity examining the distribution of color and altitudes of hazes as well as their relation. A variety of analyses is presented: a) Principal Component Analysis (PCA); b) color-altitude indices; and c) chromaticity diagrams (for a quantitative characterization of Jupiter "true" colors as they would be perceived by a human observer). PCA of the full dataset indicates that six components are required to explain the data. These components are likely related to the distribution of cloud opacity at the main cloud, the distribution of two types of hazes, two chromophores or coloring processes and the distribution of convective storms. While the distribution of a single chromophore can explain most of the color variations in the atmosphere, a second coloring agent is required to explain the brownish cyclones in the North Equatorial Belt (NEB). This second colorant could be caused by a different chromophore or by the same chromophore located in structures deeper in the atmosphere. Color indices separate different dynamical regions where cloud color and altitude are correlated from those where they are not. The Great Red Spot (GRS) appears as a well separated region in terms of its position in a global color-altitude scatter diagram and different families of vortices are examined, including the red cyclones which are located deeper in the atmosphere. Finally, a chromaticity diagram of Jupiter nearly true color images quantifies the color variations in Jupiter's clouds from the perspective of a visual observer and helps to quantify how different are the observed shades of yellow and red. The color analysis also gives additional evidence in favor of a second distinct color in the red brown cyclones of the NEB.

**Highlights**

- We explore the distribution of colors and hazes in Jupiter from Cassini images.

- Principal Component Analysis identifies two sources of colors.

- Jupiter true colors are evaluated as they would be observed by a human observer.

**\*Detailed Response to Reviewers**

We have updated the reference of Mendikoa et al. 2012, and we have done some minor grammar correction ('indexes' to 'indices').

# Spatial distribution of Jovian clouds, hazes and colors from Cassini ISS multi-spectral images

**I. Ordonez-Etxeberria (1,2), R. Hueso (1), A. Sánchez-Lavega (1),**

**S. Pérez-Hoyos (1)**

(1) Departamento de Física Aplicada I, E.T.S. Ingeniería, Universidad del País Vasco,

Alameda Urquijo s/n, 48013 Bilbao, Spain.

(2) Isaac Newton Group of Telescopes, Santa Cruz de La Palma, Canary Islands, Spain.

*Corresponding author address:*
Inaki Ordonez-Etxeberria
Dpto. Física Aplicada I,
E.T.S. Ingeniería, Universidad del País Vasco,
Alda. Urquijo s/n, 48013 Bilbao, Spain
E-mail: iordonez002@ehu.es

**Abstract**

The Cassini spacecraft made a gravity assist flyby of Jupiter in December 2000. The Imaging Science Subsystem (ISS) acquired images of the planet that covered the visual range with filters sensitive to the distribution of clouds and hazes, their altitudes and color. We use a selection of these images to build high-resolution cylindrical maps of the planet in 9 wavelengths. We explore the spatial distribution of the planet reflectivity examining the distribution of color and altitudes of hazes as well as their relation. A variety of analyses is presented: a) Principal Component Analysis (PCA); b) color-altitude indices; and c) chromaticity diagrams (for a quantitative characterization of Jupiter "true" colors as they would be perceived by a human observer). PCA of the full dataset indicates that six components are required to explain the data. These components are likely related to the distribution of cloud opacity at the main cloud, the distribution of two types of hazes, two chromophores or coloring processes and the distribution of convective storms. While the distribution of a single chromophore can explain most of the color variations in the atmosphere, a second coloring agent is required to explain the brownish cyclones in the North Equatorial Belt (NEB). This second colorant could be caused by a different chromophore or by the same chromophore located in structures deeper in the atmosphere. Color indices separate different dynamical regions where cloud color and altitude are correlated from those where they are not. The Great Red Spot (GRS) appears as a well separated region in terms of its position in a global color-altitude scatter diagram and different families of vortices are examined, including the red cyclones which are located deeper in the atmosphere. Finally, a chromaticity diagram of Jupiter nearly true color images quantifies the color variations in Jupiter's clouds from the perspective of a visual observer and helps to quantify how different are the observed shades of yellow and red. The color analysis also gives additional evidence in favor of a second distinct color in the red brown cyclones of the NEB.

# 1. Introduction

The Jovian atmosphere is covered by clouds and hazes that display significant contrast in their visible appearance. True color images of the planet reveal pale colors with brown, orange and yellow hues at different locations (Peek, 1958; Owen and Terrile, 1981; Simon-Miller et al. 2001a, 2001b). The origin of the colors is uncertain. Ammonia ice and water are colorless, and $NH_4SH$ ice acquires a bluish color when irradiated with ultraviolet light (Lebofsky and Fegley, 1976). However, trace amounts of sulfur, phosphorus and/or organic compounds could combine to form the observed colors. Candidate materials are listed by West et al. (2004), and recent research involving photolysed ammonia and acetylene in the upper troposphere has been presented by Carlson et al. (2012) and Baines et al. (2014).

The current understanding of the Jovian vertical cloud structure is summarized by West et al. (2004). On a global scale the planet is covered by: (1) A stratospheric haze which absorbs in ultraviolet, and is more abundant at the Equator and in polar latitudes (Vincent et al. 2000; Barrado-Izagirre et al., 2008); (2) An ubiquitous haze probably made of sub-micron particles extending from 100 to 500 mbar that is more dense at the Equator and over anticyclones (West et al., 2004); (3) A cloud deck that contains most of the fine-scale observable in images acquired in the visible wavelength range. These clouds are located at pressures $\sim 0.5 - 1$ bar, consistently with thermochemical models that predict ammonia to condense at these altitudes (Atreya et al. 1999; Atreya and Wong, 2005).

The altitude and opacity of the clouds and hazes can be assessed from comparison of observations in the continuum and in methane absorption bands, which in the optical range are centered at 619, 724 and 890 nm. In these wavelengths depending on altitude, the clouds or hazes reflect the photons before they are absorbed by the atmospheric methane. The hazes are also observable in reflected ultraviolet light as dark features. Polar features seen in the near-UV and in the strong methane bands (Rages et al. 1999, Vincent et al. 2000) are probably near the 100 mbar pressure level (Barrado-Izagirre et al. 2008) and higher features at 3 mbar are expected to be related to auroral processes (Pryor et al. 1991). Photons from the continuum cross along the hazes and are reflected in the cloud deck.

Most studies locate the chromophores in the upper troposphere at altitudes between 0.3 and 0.7 bar (West et al. 1986; Simon-Miller et al. 2001b; West et al. 2004). Recent works favor the location of the chromophores in the tropospheric haze (Pérez-Hoyos et al.

2009; Strycker et al. 2011) but these results are model dependent and are based on existing observations which are more sensitive to the tropospheric haze properties and it is currently unknown if the chromophores may extend deeper into the ammonia cloud.

The spatial distribution of clouds, hazes and chromophores is linked to the atmospheric dynamics. Dark reddish belts and white zones correspond to regions of cyclonic and anticyclonic vorticity and a variety of white, red and brown vortices are found with anticyclonic and cyclonic vorticities (Ingersoll et al. 2004; Simon et al., 2015). The traditional view is that belts and zones represent regions of large-scale downwelling and upwelling respectively (Ingersoll et al. 2004). Color differences could arise either because the clouds in the belts could be covered by a reddening agent possibly produced by photochemistry, or alternatively because the belts are deeper in the atmosphere, allowing sublimation of the overlying ammonia ice rime on a core of redder material (West et al. 1986, Kuehn and Beebe 1993). Neither of these ideas has been verified, in part because of a lack of detailed information about vertical motions, and in part because of the complex coloration of the vortices and other dynamical features. Active convective storms tend to be white, implying that they are formed by fresh ices uncontaminated with colored agents. These clouds constitute the only locations of the planet where the spectral signature of ammonia ice has been found. However, this signature vanishes in time-scales of a few days indicating that some unknown coating compound, possibly of hydrocarbon origin (Kalogerakis et al. 2008), or photochemical process tan the fresh aerosols quickly (Baines et al. 2002).

Quantitative analyses of Jupiter colors are available in the literature since the Voyagers flybys (Owen and Terrile, 1981; West et al. 1986; Thompson et al. 1990). Simon-Miller et al. (2001a) performed a Principal Component Analysis (PCA) of Hubble Space Telescope (HST) Jupiter observations avoiding filters with contributions from methane absorption bands. Therefore, they investigated only the main visible cloud level without contributions from its vertical cloud structure. The continuum range of Jupiter colors could be explained by the spatial distribution of only two chromophores, although the conclusions regarding the second chromophore were weak. A similar PCA analysis of HST images of oval BA also suggested the requirement of a second chromophore in the GRS, but the analysis was focused on a small area of the planet (Simon-Miller et al. 2006). These works were expanded by Strycker et al. (2011) who made a stronger case for the existence of a second chromophore but analyzed only a portion of the equatorial and low latitudes of the planet. Other authors have used PCA to summarize the spectral information in hyperspectral images from the NIMS and ISS instruments on Galileo (Dyudina et al. 2001), or have used PCA results in radiative transfer models of the clouds (Irwin and Dyudina, 2002).

Changes in coloration that occur in Jovian anticyclones have been particularly interesting (Sánchez-Lavega et al. 2013). The anticyclone BA, formed after the fusion of three long-lived White Ovals (Sánchez-Lavega et al. 1999, 2001), changed from white to red in 2006 (Simon-Miller et al. 2006). Radiative transfer models of HST data by Pérez-Hoyos et al. (2009) determined that the change occurred in the 250–450 nm wavelength range and favored a subtle change in the tropospheric particle properties. The color change was not accompanied by a change in the cloud morphology, dynamical interactions with other features, changes in its dynamics or in its temperature (García-Melendo et al., 2009; Hueso et al. 2009; Sussman et al. 2010; Wong et al. 2011; Cheng et al. 2008). Other interesting vortices were the so-called Red Oval, a strongly red colored anticyclone that was engulfed by the GRS around June 2008 without altering the color of the GRS (Strycker et al. 2011; Sánchez-Lavega et al. 2013), and a Red Cyclone visible in Jupiter in 1994 and 1995 that was intensely colored at a latitude where white features are generally found (Simon et al. 2015).

Large-scale color changes also occur in certain belts and zones in time-scales of a few months to a year. The South Equatorial Belt (SEB) of Jupiter alternates at visible wavelengths from a belt-like band to a short-lived zone-like aspect (Sánchez-Lavega and Gómez, 1996). These changes are correlated with changes in the atmospheric temperature and aerosol opacity (Fletcher et al. 2011) and occur within the cloud decks in the convective troposphere (Fletcher et al. 2011; Pérez-Hoyos et al. 2012). Disturbances in the North Temperate Belt produced by outbursts of convective activity also result in large-scale color changes (García-Melendo et al., 2005; Sánchez-Lavega et al., 2008).

Other questions related to Jupiter colors and the phenomena that intervene remain largely unanswered. Here we explore the global distribution of colors and altitudes in Jupiter clouds from multi-spectral images of the planet acquired by the Cassini ISS during its Jupiter flyby. Section 2 describes the observations and data processing. Section 3 presents analyses of the data using PCA. Section 4 analyzes color and cloud altitudes using color indices in parallel to recent work presented by Sánchez-Lavega et al. (2013). Section 5 contains a quantitative description of true colors in Jupiter's atmosphere based on a standard chromaticity diagram. We discuss our results in section 6.

## 2. Cassini ISS data and basic processing

The Cassini Jupiter flyby provided high-resolution observations of the Jupiter system with the Imaging Science Subsystem (ISS). The ISS consists of two CCD cameras with a narrow (NAC) and wide field of view (WAC) and a large selection of filters. Both CCDs are of the same type and their sensitivity extends from the ultraviolet to the near infrared (200-1050 nm) although the WAC misses the 200-380 nm wavelength range, because of its refracting optics. Details about the ISS instrument, its performance, and the filters available are given by Porco et al. (2004). We downloaded Cassini ISS images from the NASA PDS as well as the SPICE kernels required to navigate these images.

*2.1. Data selection.*

Our aim was to build nearly full cylindrical maps of the planet with the smallest possible number of images, the highest spatial resolution and the largest number of filters. We were interested in observations acquired with narrow band filters at the weak (MT1, centered at 619 nm), intermediate (MT2, 727nm) and strong (MT3, 890nm) methane absorption bands and their adjacent wavelengths (CB1 – 619 nm; CB2 – 750 nm; CB3 – 938 nm) to explore the spatial distribution of tropospheric clouds and upper hazes. These images were complemented by observations acquired in the ultraviolet (UV1 – 258 nm). We also analyzed images sensitive to color comparing the previous images with observations acquired with the broadband BL1 (451 nm) and GRN (568 nm) filters. Unfortunately, there were no observations in the broadband RED filter in this time series.

We selected 54 consecutive images acquired on 14 November 2000 by the Cassini ISS NAC that fulfilled these requirements. The observations were obtained from a distance to the planet center of 44 million kilometers and had a spatial resolution of ~262 km per pixel at the sub-spacecraft point. The UV1 and MT3 images were binned onboard the spacecraft to improve the signal to noise ratio resulting into a spatial resolution of ~525 km per pixel. Although the Cassini flyby obtained many subsequent observations with higher spatial resolution their analysis in terms of full maps requires the navigation and projection of a significantly larger number of images. Figure 1 shows examples of the images. Figure 2 shows the filter properties compared with a full disk spectrum of the planet. Details about the specific images are provided in the Appendix.

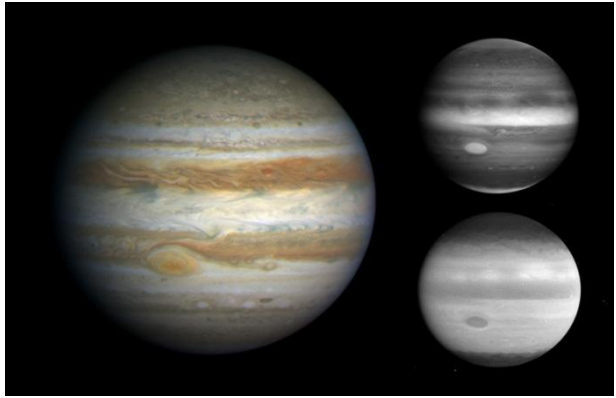

**Figure 1:** Synthetic color image from CB2, GRN and BL1 images (left) compared with MT3 (upper right) and UV1 images (lower right). All images have been contrast stretched to improve visibility of the atmospheric details.

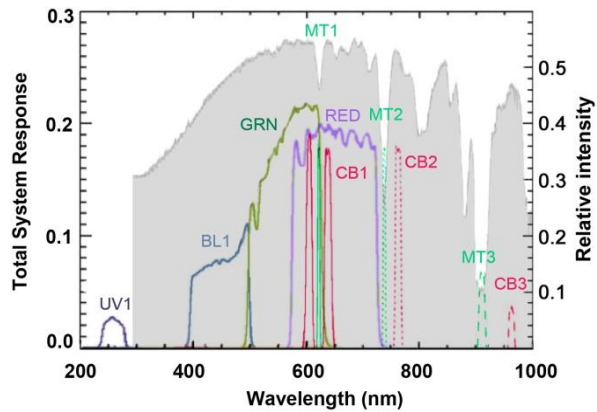

**Figure 2:** Total system response for Cassini NAC filters used in this study (left axis). Data is from Porco et al. (2004). Note the double-lobe structure of the CB1 filter. The data is superimposed over a reflection spectrum of the planet (right axis) from data published by Karkoschka (1998).

*2.2. Image navigation and maps building*

Images were flux calibrated using the Cassini Imaging Science Subsystem CALibration (CISSCAL) software as described in West et al. (2010). The calibrated images were loaded into the PLIA software (Hueso et al. 2010) which makes use of SPICE kernels to navigate the images and obtain all the required geometry information including the illumination scattering angles, $\mu$=cos $e$ and $\mu_0$=cos $i$, where $e$ is the emergence angle and $i$ is the incidence angle at each pixel. The image navigation was checked and corrected by comparing the planet limb with the limb computed from the SPICE kernels. We estimate the final image navigation to be accurate to a pixel value. We then applied a Lambert function to correct limb-darkening effects.

$$\frac{(I/F)}{(I/F)_0} = \mu_0. \tag{1}$$

Here $(I/F)$ is the observed atmospheric reflectivity and $(I/F)_0$ the reflectivity at nadir view with zenithal illumination. This correction was efficient for all wavelengths except in the UV1 filter due to the strong Rayleigh scattering close to the limb. These images were corrected using a Minnaert function (Minnaert, 1941) which produced better results. The Minnaert correction is defined as

$$\frac{(I/F)}{(I/F)_0} = \mu_0^k \mu^{k-1}, \tag{2}$$

where $k$ is a fixed index that defines the Minnaert correction and is experimentally set to 0.61 for the UV1 filter ($k$=1 is equivalent to the Lambert law).

All the images were cylindrically projected into longitude (system II) and planetocentric latitude maps of 0.1 degree per pixel, corresponding to a spatial resolution of 124.8 km at the equator. This spatial resolution oversamples the original images. While this has no consequences for the current analysis, it improves the visibility of some faint details and the comparison with other maps of Jupiter. We combined images acquired sequentially over a planetary rotation to build full longitudinal cylindrical maps of the planet in each filter. The pixel values at overlapping regions between two images were computed by weighted linear interpolation (Barrado-Izagirre et al., 2009). Figure 3 shows examples of the cylindrical maps.

*2.3. Synthetic Red*

In order to compare PCA color analysis from this work with previous analysis and explore the range of true colors available in the planet we built a synthetic red image that could be comparable to an image obtained by the RED filter on Cassini ISS. Since Jupiter's spectrum is very flat at red wavelengths and images in CB1, CB2 and RED only vary subtly, any of the narrow filters CB1 or CB2 could be used as first-order representative of the red information and are generally used for enhanced color contrast representations of Jupiter colors. However, a better approach is to combine both filters to build a synthetic RED image that can be used to construct a nearly true color map of the planet. We combine CB1 and CB2 calibrated maps into a single red map whose effective wavelength and radiant flux is the same as the effective wavelength and integrated response to solar light that would have been obtained by the RED filter inside the ISS filter wheel. Formally the image combination can be obtained by solving the following system of equations:

$$F_R = aF_{CB_1} + bF_{CB_2},$$
$$\lambda_R = a\lambda_{CB_1}\frac{F_{CB_1}}{F_R} + b\lambda_{CB_2}\frac{F_{CB_2}}{F_R}. \quad (3)$$

Here $a$ and $b$ are non-dimensional parameters, $\lambda_R = 650$ nm, $\lambda_{CB1} = 635$ nm and $\lambda_{CB2} = 752$ nm are the effective wavelengths of a RED, CB1 and CB2 images and $F$ stands for the flux of the Sun convolved by the spectral transmisivity for each of the three filters. The solution to this system is given by $a=5.74$ and $b=2.27$, which translates into the following image combination.

$$RED^* = a\frac{F_{CB_1}}{F_R}CB_1 + b\frac{F_{CB_2}}{F_R}CB_2 = 0.8718CB_1 + 0.1282CB_2 \quad (4)$$

With this image in hand, a photometric RED* GRN BL1 image constitutes a nearly true color representation of color in Jupiter. The result is shown in the first panel in Figure 3 which also includes maps in the MT3 and UV1 filters.

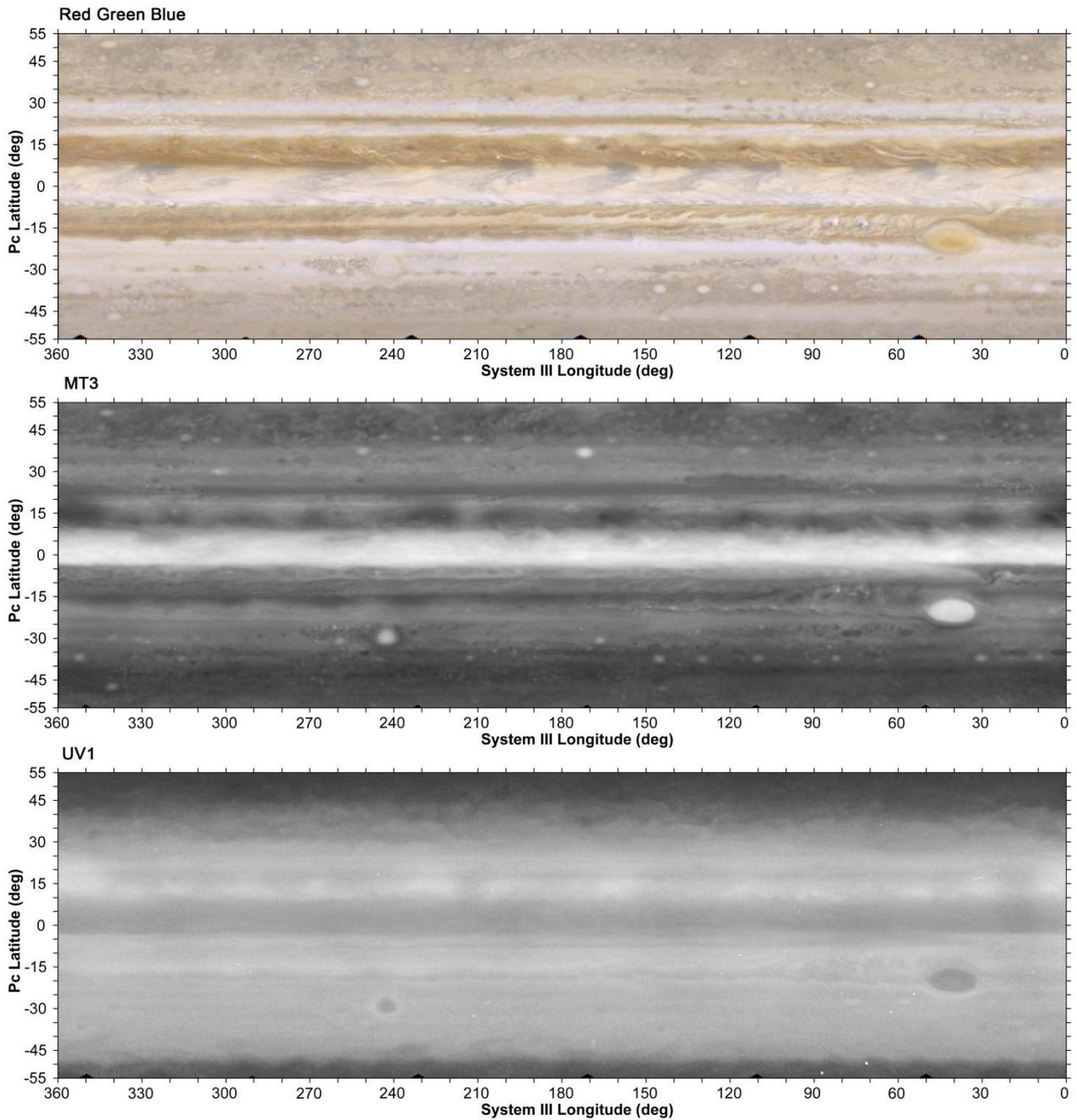

**Figure 3:** Jupiter cylindrical maps in different wavelengths. Top: Nearly true color map combining RED*, GRN and BL1 maps; Middle: Map in the strong methane absorption band at 890 nm (MT3 filter); Bottom: Map at 258 nm in the ultraviolet range (UV1 filter).

## 3. Principal Component Analysis

*3.1 PCA Basics*

PCA is a standard technique in multivariate data analysis with wide applications in image processing (Murtagh and Heck, 1987) and remote sensing (Ready and Wintz, 1973).

PCA is a linear transformation of statistical nature that represents the data in a new coordinate system in which basis vectors formed by a linear combination of the original variables (i.e. the original images) follow modes of greatest variance in the data. The main drawback to the use of PCA is the difficulties that arise in the interpretation of the individual PCs. However, PCA of multi-spectral images is regularly used to search for patterns in multi-spectral data of the Earth towards feature classification (Richards and Jia, 2006).

The result of a PCA of a set of *n* images $Im_j$ of the same object acquired with different filters (where the subscript *j* identifies each image) is a set of *n* principal components $PC_i$ (or eigenvectors here numbered with a subscript *i*) defined by a set of *n* x n weights, $w_{ij}$, that define each $PC_i$ as a linear combination of the original images. The analysis also provides the variance associated to each principal component as measured by the eigenvalue of each PC and, for convenience, PCs are ordered from higher to lower variance and defined by weighted means of the original images.

$$PC_i = \sum_{j=i}^{n} w_{ij} \, \text{Im}_j \qquad (5)$$

Since the choice of signs for weights $w_{ij}$ is arbitrary, the interpretation of PCs can sometimes be aided by reversing them.

*3.2. Color distribution from PCA*

The synthetic RED*, GRN and BL1 images contain the color information in true color images of the planet and compare well in terms of spatial resolution with HST observations analyzed by Simon-Miller et al. (2001a; 2006). However these observations are better suited to study the distribution of visual colors in Jupiter, since all filters are wide and approximate better Jupiter true-colors than previous analyses. Later PCA studies of HST images have focused on the blue part of the spectrum and have been limited to portions of the equator and low latitudes (Strycker et al. 2011). The three PCs obtained from our set of maps are shown in Figure 4A. The first component accounts for 94.2±0.4% of the variance and describes the overall brightness variations. The second component contains 5.5±0.2% of the variance and a third component describes only a 0.3±0.2% of the variance. Each PC is obtained from the image weights summarized in Table 1. Errors in these numbers constitute upper limits and come from extensive tests modifying the images in two ways. Firstly, we introduced a 1% noise level for each pixel, representative of an assumed photometric precision of 1%, and, secondly, we considered co-registration errors from possible navigation errors redoing the analysis several times shifting each map one pixel in different directions. The analysis of these experiments results in the error bars given above. All weights in Table 1 can be computed to the third decimal with errors on the order of $4\times10^{-4}$. PCs with information content at levels below 0.2% represent the noise floor limit caused by navigation and coregistration errors which dominate the PCA errors.

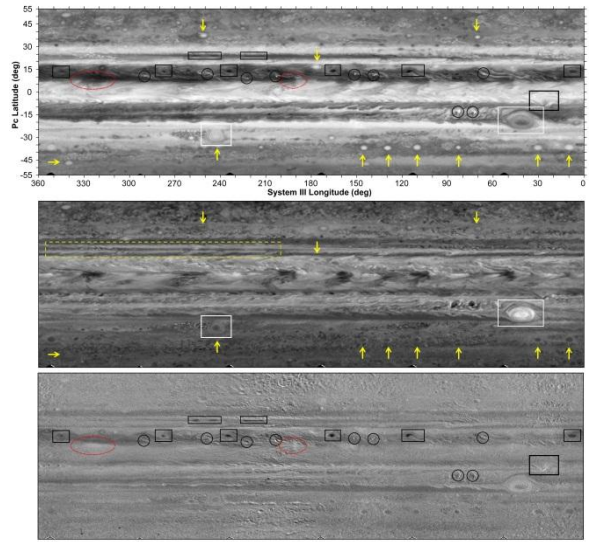

**Figure 4A:** Principal Components 1, 2 and 3 from top to bottom from analysis of RED*, GRN and BL1 cylindrical maps. All images have been contrast-stretched to show the underlying structure. Highlighted regions are discussed in the text. The upper panel contains shared longitudes and latitudes between the PC maps.

Figure 4A highlights several regions. The North Equatorial Belt (NEB) at planetocentric latitudes ~ 7–17º) stands out in PC2 as the brightest latitudinal band of the planet, and the GRS is the brightest area while it is relatively dark in the first PC. This contrasts very strongly with the appearance of oval BA (planetocentric latitude -29º and system III longitude 243), also highlighted with a large white box, bright in PC1 and dark in PC2. White anticyclones are pointed with vertical arrows; they are bright in PC1 and have negligible contrast in PC2. The North Temperate Belt (NTB) at planetocentric latitudes 20-25º stands out as a bright feature in PC2 and has a longitudinal structure: Its western side, highlighted with a dashed box, presents a double branch in PC2 and a single branch in the Eastern side. This double structure in this region is not apparent in PC1 or PC3. The darkest features in PC3 are cyclonic ovals (highlighted in PC1 and PC3 with small black boxes) similar to the barges sometimes observed in the planet. The brightest features in PC3 are convective storms, highlighted in PC1 and PC3 with black circles. Other interesting features are the equatorial dark projections (two of the most prominent are highlighted with dotted ellipses). They are dark in PC1 and bright in PC3. Another bright feature in PC3 is the South Equatorial Disturbance, SED (Garcia-Melendo et al. 2011) , highlighted with a black box at planetocentric latitude -5º and system III longitude 25º North East of the GRS.

Figure 5A shows the amplitudes of the Red*, GRN and BL1 image that result in the three PCs. Together with figure 4A it can be used to interpret the information shown in each PC. Our interpretation is the following. The first PC is related to the overall cloud reflectance and is a weighted average of the three images. Bright features in the second PC correlate well with the red regions in the planet and dark features in PC2 correspond to white features in color images of the planet. PC2 is formed by a positive contribution of the red part of the images and a negative contribution from the blue part of the images (Figure 5A and Table 1). Therefore, PC2 can be interpreted as a first order map of the red color that tints the cyclonic belts including the NEZ, the GRS and other smaller ovals (every feature that is brighter than its surroundings in PC2 is redder than its surroundings in the original nearly true color map). However small, the third PC is not absent of information and the spatial structure visible in PC3 implies that a third component is required to explain the data. PC3 has extreme brightness values at well identified regions: convective storms, the very slightly greenish SED, and most notably, the small dark reddish cyclones similar to the Jovian barges. The different behavior of red cyclones and barges has been remarked recently in a detailed analysis of one rare and particularly red cyclone in the planet visible in 1994 and 1995 (Simon et al. 2015).

Two interpretations of the information in the third PC are possible: (A) A single chromophore explains most of the shades of color in Jupiter's clouds and a second chromophore is required to explain the color in certain regions where PC3 has higher or lower brightness values. (B) A single chromophore explains all the colors but the cloud particles have different optical properties in specific locations, perhaps by being deeper in the atmosphere. In both cases we can conclude that two coloring process are present in the Jovian clouds. Additional support to two sources of color is provided in the later sections as well as arguments in favor and against the two chromophores interpretation. Therefore, dark compact features in PC3 will be interpreted as regions where a second coloring process is present. We note that the third PC, associated to the second coloring process, is an important part of the GRN image and accounts for 0.8% of the variance associated to the GRN filtered image when doing the inverse PC transformation (i.e. when building the RED*, GRN and BL1 images from the three PCs).

Variances associated to each PC in this work agree very well with numbers obtained from similar analyses of colors in Jupiter from HST observations with color sensing filters (Simon-Miller et al. 2001a; 2006). A recent analysis of HST multi-spectral images (Strycker et al. 2011) also agrees with the numbers obtained here and deepens into the interpretation of the data as an evidence of a second chromophore agent. We go further to indicate the regions where this second chromophore should have concentrations much larger or smaller than in the rest of the atmosphere. We also point out that differences between the first chromophore and the second chromophore or coloring process concentrate in the green region of the spectrum.

*3.3. Color distribution from an extended PCA*

An extended analysis was run for five images: BL1, GRN, CB1, CB2 and CB3. The information or image variance is now distributed in 5 PCs with information contents of 89.5±0.3%, 8.8±0.2%, 1.0±0.2%, 0.4±0.1% and 0.2±0.1% for increasing PCs. Errors associated to these numbers are estimated as in the previous case and slightly smaller due to the use of a higher number of images. The 5 PCs are shown in Figure 4B, which is highlighted similarly to Figure 4A. The contributions to each PC are summarized in Table 1 and Figure 5B. A visual comparison of PCs from this case with the previous one shows that PC1 and PC2 are essentially the same as previously, and that PC3 retains very similar structures and contrast. The larger number of images used results in improved visibility of the structures in PCs 2 and 3, with noise now transported to PCs 4 and 5. Figure 5B and 5A show also similarities in terms of the composition of each PC from the original images. PC4 retains a certain degree of spatial structure and may isolate anomalous regions in terms of colors. Convective storms and small dark cyclones are now well separated in PC4. PC5 (containing only a 0.2% of the total variance) can be fully explained from the noise produced by image navigation errors resulting in small co-registration defects.

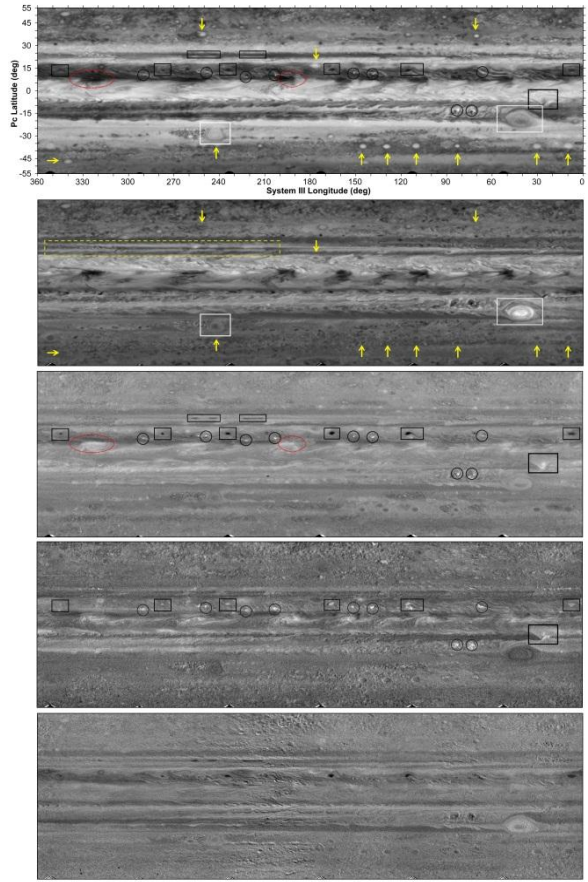

**Figure 4B:** Principal Components 1 to 5 ordered from top to bottom obtained from analysis of CB1, CB2, CB3, GRN and BL1 cylindrical maps. All images have been contrast-stretched to show the underlying structure. Highlighted regions are discussed in the text. The upper panel contains shared longitudes and latitudes between the PC maps.

Again, the brown reddish cyclones and the convective storms in the wake of the GRS and in the NEB can be considered as anomalous regions with different spectral properties to the rest of the planet. The convective regions are expected to be "anomalous" on the basis of the fresh ("bright") material produced at the storm and the brown reddish cyclones might be locations of chromophores, something already concluded from Voyager images of the equivalent brown barges observed at the epoch (West et al. 1986).

*3.4. Global PCA analysis.*

We ran an extended analysis over the 9 maps. The results are more complex due to the added information and the correlations between many of the images. Figure 4C shows the first six PCs also defined in Table 1. The image amplitudes for the first 5 PCs are represented in Figure 5C. At least 6 components are required to adequately represent the data. However the PCs continue to be highly correlated hindering a good separation of the underlying causes of the spatial variations.

Again, PC1 represents the overall variance of the data and contains information from the cloud and hazes distribution as well as color. PC1 contains a 73.5±0.3% of the total variance of the data. PC2, accounting to a 19.5±0.2% of the total variance, contains mixed information from the UV1 and BL1 images with negligible contributions from other images (Figure 5D and Table 1). PC2 mixes the color information from the BL1 map with the anticorrelated information in UV1. The morphology in PC3 correlates with the spatial distribution of color and PC3 is visually similar to previous color maps on the previous simplified color analysis. It accounts to a 5.0±0.2% of the total variance of the data. PC4 groups pixels that are anomalous in terms of their color, and acquires anomalous bright or dark values at the dark-reddish cyclones and the convective storms containing about 0.7±0.1% of the total variance. PC5 correlates well with the distribution of upper hazes, and contains a 0.7±0.1% of the total variance. Finally, the sixth PC, representing a total variance of only 0.3±0.1% of the original data, continues to have spatial information at particular locations, notably at regions with active moist convection. The next 3 components contain less than 0.4% of the data variance and, although they have some spatial structure, they probably only represent the effects of noise in the image composition from co-registration of the images. The 6 components contain mixed information from at least six different sources. We tentatively interpret these sources as clouds, hazes of two types (from their different contribution to UV and MT3 images), two unknown sources of color and convective storms.

From the ensemble of these analyses we conclude that PCA is not able to adequately separate the information from all these sources and in particular is not able to separate the effects of the lower clouds and upper hazes. However, PCA provides reasonable results in characterizing the spatial distribution of colors from two different sources and signals the location of interesting features.

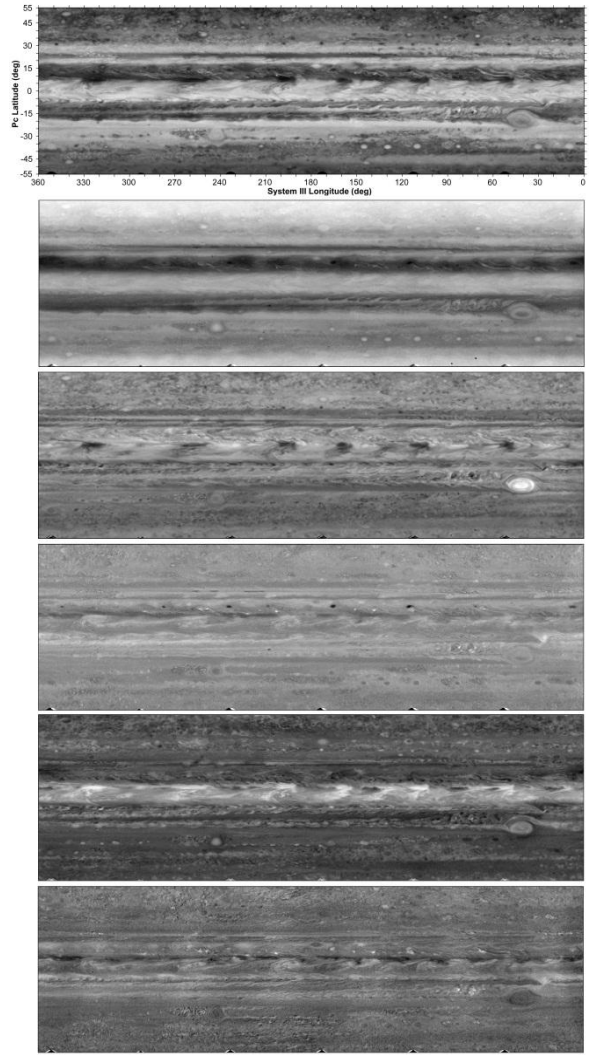

**Figure 4C:** First six principal components ordered from top to bottom obtained from analysis of the 9 cylindrical maps. All images have been contrast-stretched to better show the underlying structure. The upper panel contains shared longitudes and latitudes between the PC maps.

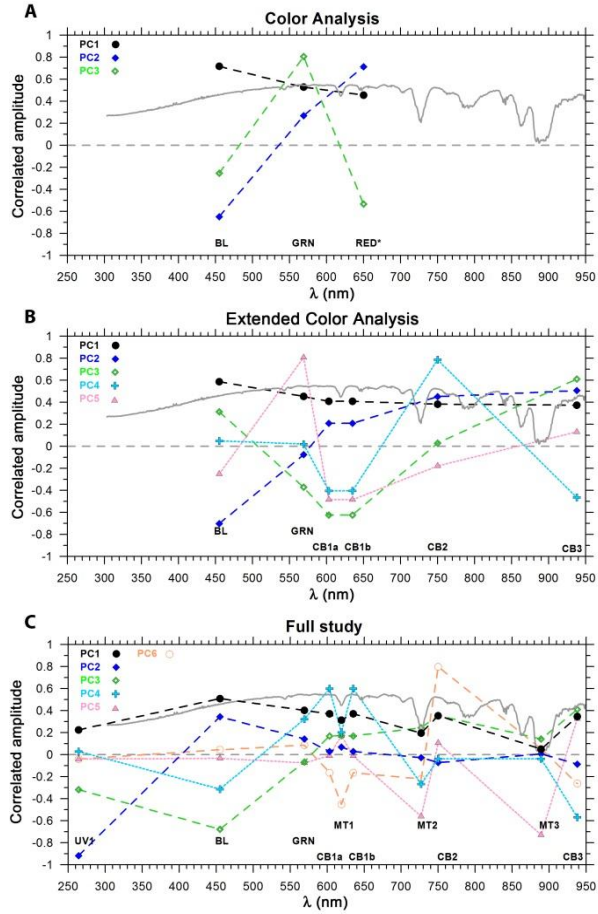

**Figure 5:** Correlated amplitudes that define Principal Components from the four analyses (from top to bottom: color [A], extended color [B], and global analysis [C]). The data is superimposed over a reflectivity spectrum of Jupiter (Karkoschka, 1998). Positive values indicate positive correlation of the individual images with each PC, negative coefficients indicate anticorrelation. Note how PCs containing information about the color distribution from different analysis (PC2 in cases A and B and PC3 in C) share the same spectral behavior.

**Table 1 - Principal component coefficients**

| PC | Filters | | | | | Variance | Comment |
|---|---|---|---|---|---|---|---|
| | BL1 | GRN | RED* | | | | |
| **Basic color study** | | | | | | | |
| 1 | 0.717 | 0.528 | 0.455 | | | 94,2±0.4% | Cloud |
| 2 | -0.649 | 0.269 | 0.712 | | | 5.5±0.2% | Color |
| 3 | -0.253 | 0.806 | -0.535 | | | 0.3±0.2% | AR |
| **Advanced color distribution** | | | | | | | |
| | BL1 | GRN | CB1 | CB2 | CB3 | | |
| 1 | 0.587 | 0.452 | 0.408 | 0.382 | 0.373 | 89,5±0.3% | Cloud |
| 2 | 0.702 | 0.076 | -0.209 | -0.451 | -0.505 | 8,8±0.2% | Color |
| 3 | -0.313 | 0.372 | 0.624 | -0.029 | -0.611 | 1.0±0.2% | AR |
| 4 | 0.049 | 0.020 | -0.405 | 0.786 | -0.464 | 0.4±0.1% | Noise |
| 5 | -0.251 | 0.807 | -0.486 | -0.178 | 0.131 | 0.2±0.1% | Noise |

|   | *Full study* | | | | | | | | | | |
|---|------|------|------|------|------|------|------|------|------|-----------|--------------|
|   | UV | BL1 | GRN | MT1 | CB1 | MT2 | CB2 | MT3 | CB3 | | |
| 1 | 0.226 | 0.511 | 0.403 | 0.314 | 0.372 | 0.196 | 0.355 | 0.051 | 0.348 | 73.5±0.3% | Cloud & Haze |
| 2 | -0.917 | 0.345 | 0.144 | 0.070 | 0.030 | -0.026 | -0.071 | 0.008 | -0.085 | 19.5±0.2% | Color & Haze |
| 3 | -0.317 | -0.677 | -0.065 | 0.174 | 0.171 | 0.246 | 0.356 | 0.143 | 0.412 | 4.9±0.2% | Color |
| 4 | 0.031 | -0.312 | 0.322 | 0.203 | 0.600 | -0.266 | -0.037 | -0.036 | -0.569 | 0.72±0.2% | AR |
| 5 | -0.040 | -0.033 | -0.074 | 0.173 | -0.009 | -0.560 | 0.107 | -0.728 | 0.327 | 0.69±0.2% | Haze |
| 6 | -0.040 | 0.045 | 0.088 | -0.450 | -0.163 | -0.220 | 0.799 | 0.055 | -0.262 | 0.30±0.1% | AR |
| 7 | -0.034 | 0.103 | -0.242 | -0.588 | 0.589 | -0.249 | -0.164 | 0.191 | 0.332 | 0.19±0.1% | Noise |
| 8 | 0.006 | -0.224 | 0.793 | -0.309 | -0.249 | -0.142 | -0.251 | 0.057 | 0.278 | 0.15±0.1% | Noise |
| 9 | 0.033 | 0.036 | -0.092 | 0.388 | -0.187 | -0.621 | 0.015 | 0.635 | 0.121 | 0.05±0.1% | Noise |

Note: AR stands for anomalous regions that dominate the principal component image and correspond to particular regions such as cyclonic and convective storms.

*3.5. PC interpretation in terms of simple altimetry.*

In order to provide some physical insight into the previous interpretation we combine images in CB2 and MT3 filters in terms of a simple single reflecting layer model following Mendikoa et al. (2012). Briefly, if we compare the reflected light in two similar wavelengths, with one of them subject to additional absorption by atmospheric methane, neglecting differences in Rayleigh scattering between them, we can compute the excess optical depth of the atmosphere in the methane sensitive image, $\tau_g$, using a single reflective layer model (Chandrasekhar 1960; Hansen and Travis 1974).

$$\left(\frac{I}{F}\right)_{MT3} = \left(\frac{I}{F}\right)_{CB2} e^{-\tau_g\left(\frac{1}{\mu_0}+\frac{1}{\mu}\right)} \quad (6)$$

where we follow the same notation as in section 2.2. Note that this comparison holds for both the original images and our Lambert-corrected maps from eq. (1) since each pixel in the maps in both wavelengths has the same multiplicative correction.

Equation 6 can be inverted to compute the added optical depth due to methane at 890 nm above the upper cloud observed in the CB2 map. For this purpose, we computed the average value of the scattering angles $\mu$ and $\mu_0$ to adequately represent the geometry of the observation.

$$\left(\frac{I}{F}\right)_{MT3} \simeq \left(\frac{I}{F}\right)_{CB2} e^{-\tau_g\left\langle\frac{1}{\mu_0}+\frac{1}{\mu}\right\rangle} = \left(\frac{I}{F}\right)_{CB2} e^{-3.2\tau_g}. \quad (7)$$

Where < > denotes the mean value over the map and variations over the mean of the geometric factor are less than 10%. The inverse transformation results in:

$$\tau_g = -\frac{1}{3.2}\ln\frac{(I/F)_{MT3}}{(I/F)_{CB2}}. \qquad (8)$$

This expression can be evaluated in every pixel producing a global map of methane optical depth at nadir viewing from the top of the atmosphere to the reflecting layer. Results are shown in Figure 6.

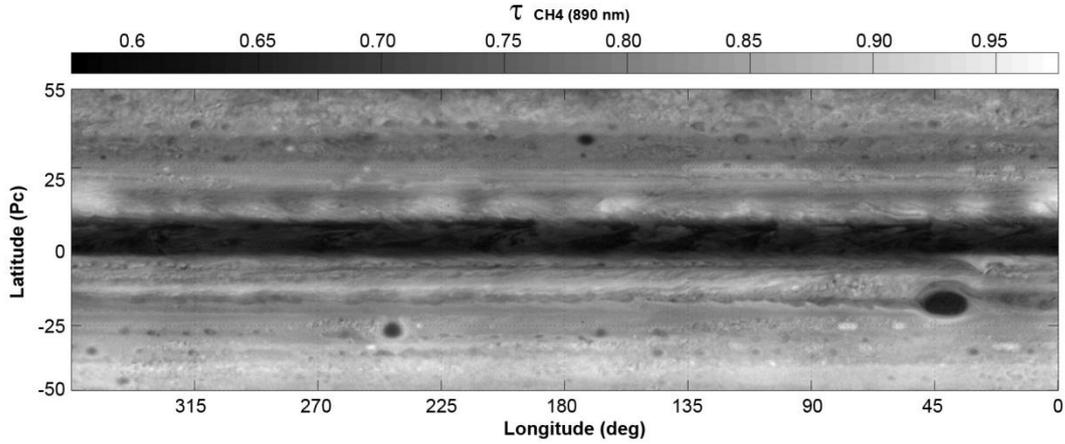

**Figure 6:** Methane optical depth from the Reflecting Layer Model. Dark structures represent high atmospheric layers clouds and bright features represent deeper clouds. Since we have used corrected limb-darkening images and only two wavelengths the results should not be overemphasized. The GRS and NEB appear as the regions covered by the highest clouds and lowest methane absorption as well as one prominent red anticyclone in the NTB. Oval BA has cloud tops slightly below the level reached at the GRS.

The resulting map of methane optical depth at 890 nm compares very well with those PCs previously identified with the structure of the upper hazes. Figure 7 shows the correlation between this map and PC5 from the global analysis supporting the interpretation that PC5 is related with cloud top altitude. The figure compares brightness values in the PC5 with the methane optical depth at each point as a scatter plot. The lower part of the scatter plot correspond to the regions covered by high hazes such as the equatorial region, the GRS and oval BA. The rest of the scatter plot is distributed uniformly over the image. A linear fit between values of optical depth and PC5 brightness has a coefficient of determination $R^2$=0.764.

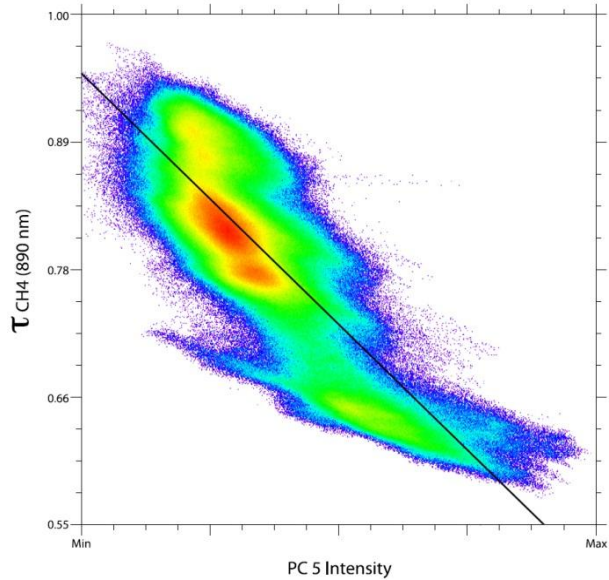

**Figure 7:** Pixel by pixel comparison of the methane optical depth from the Reflecting Layer Model with PC5 brightness from the global PCA in section 3.5. Color is used to represent the density of pixels; red corresponds to high pixel densities, blue corresponds to a low pixel density in the scatter diagram. The solid line represents a linear fit to the data.

We conclude that the morphology present in PC5 from our global analysis is mainly caused by the spatial variations of methane absorption. However, the information is partially mixed with the underlying cloud structure and cannot be used to separate the contribution of the hazes to the different images as it was originally intended. Another noteworthy characteristic is that, while red cyclones are not easily distinguishable in the methane optical depth map (Figure 6), they are in fact located within the regions of highest methane optical depth and lowest cloud top. These aspects are inconclusive in terms of the discussion of their particular color as a consequence of a second chromophore or as a consequence of a lower cloud top altitude.

**4. Color indices**

Color diagrams of Jupiter images show a continuum distribution of colors (Simon-Miller et al. 2001a) whose relation with different regions of the planet is difficult to interpret. A better technique to combine information from different filters is to build image ratio indices able to separate different regions of the planet in a single diagram. Sánchez-Lavega et al. (2013) analyzed the colors and altitudes of several vortices in Jupiter from HST images and proposed two image ratio indices based on four wavelengths to build a color altitude diagram where to classify them. Each structure (vortex, convective storms, belt or zone) accounted to a single point in a scatter diagram of Jovian features which differed in color and cloud-top altitude. de Pater et al. (2010b) and Wong et al. (2011) created color distributions from the HST blue (F435N) and red (F658N) filters with similar results. We follow the approach by Sánchez-Lavega et al. (2013) to define a Color Index (CI) and an Altitude Opacity Index (AOI) using the wavelengths closest to those used by them

$$CI = \frac{I/F(BL_1)}{I/F(CB_2)} \quad (9)$$

$$AOI = \frac{I/F(MT_3)}{I/F(UV_1)} \quad (10)$$

These two indices have the advantage of comprising most of the spatial variability in the data combining the four images that are less correlated containing the highest amount of information. The color indices are referenced to the neutral color characteristic of the white South Tropical Zone (STrZ).

We improved previous works running a pixel by pixel analysis finding the color index for each pixel in the maps. Results are shown in terms of an AOI-CI scatter plot (Figure 8) which is directly comparable to previous works based on HST data but highly expanded incorporating information from all features in the maps. Red features have low CI values that can go as low as 0.5 and white features have CI values of 1.0 to 1.1. High altitude features are bright in MT3 or dark in UV1 and have high AOI values.

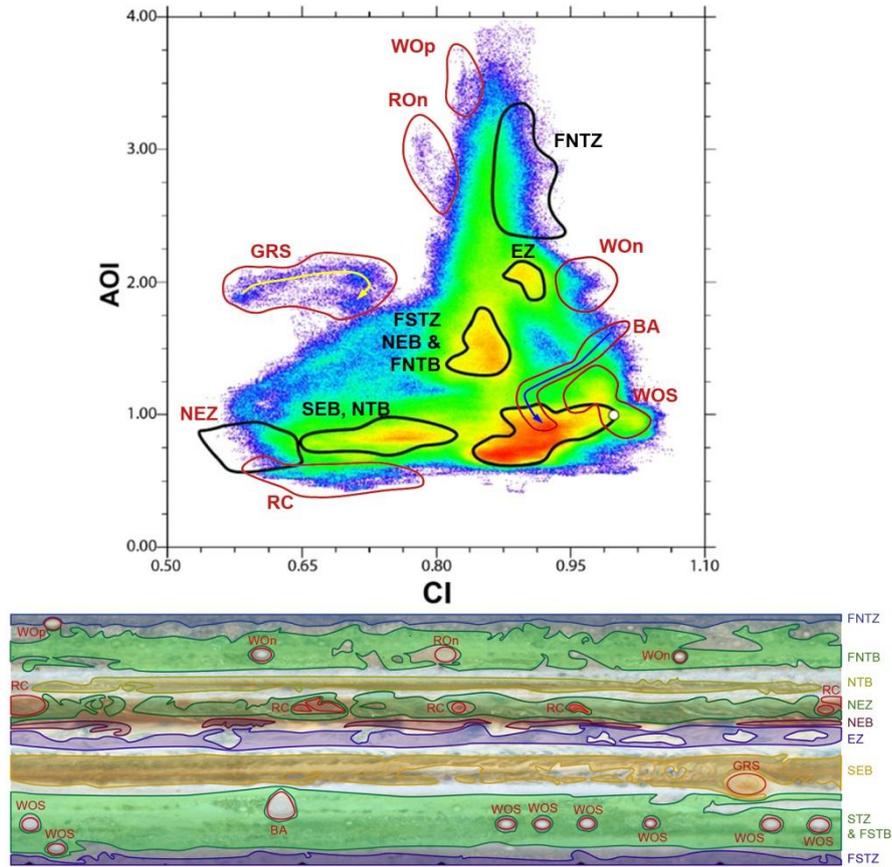

**Figure 8:** Scatter plot of color and altitude opacity indices (top) and Jupiter map (bottom). The reference region at the STrZ is indicated with a white circle. Color is used to represent the density of pixels sharing the same behavior; red regions correspond to a high abundance of pixels with the same AOI and CI values while blue regions correspond to regions less populated in the color diagram. Arrows inside the GRS and oval BA scatter plot start at the center of the oval and proceed to the outer region of the ovals. The lower panel presents an identification of the different structures. Acronyms are: FNTZ (Far North Temperate Zone), FNTB (Far North Temperate Belt), NTB (North Temperate Belt), NEZ (North Equatorial Zone), NEB (North Equatorial Belt), EZ (Equatorial Zone), SEB (South Equatorial Belt), STZ (South Temperate Zone), FSTB (Far South Temperate Belt), FSTZ (Far South Temperate Zone), WOp (White Oval Polar), WON (White Ovals North), ROn (Red Oval North), RC (Red Cyclones), GRS (Great Red Spot), BA (oval BA), WOS (White Ovals South).

The scatter plot readily identifies intrinsically different structures in terms of altitude or color. The GRS appears as a well detached structure with red hues (low CI value) and is covered by high altitude clouds or hazes (high values of the AOI). This contrasts with the rest of the Jovian atmosphere that extends over the scatter plot without sharp changes between different regions. Figure 8 shows the scatter plot together with a cylindrical map of the planet with bands, zones and vortices identified. Some features are easily interpreted. For instance, the central upper branch in the diagram corresponds to the contribution of the sub-polar hazes which are moderately bright in MT3 but very dark in UV images. The upper hazes cover much of the northern high latitudes and there is a steady gradient of the AOI from the North mid-latitudes to the high-latitudes. In terms of color, the North and South Equatorial Belts (NEB and SEB) and the North Temperate Belt (NTB) dominate the red lower part of the diagram presenting low content of upper hazes. These regions dominate the "red" like appearance of the planet. Several families of vortices are observed in the map with the main difference between them found in their position with respect to the CI axis. The reddest regions are concentrated in the GRS and the cyclonic ovals in the NEB with also low content of upper hazes. These cyclonic regions were identified as "anomalous" in color in terms of the PCA presented before. They could be the possible location of a second chromophore, but alternatively the absence of upper hazes may allow the lower atmosphere to be better observed providing better contrasted colors for the same chromophore.

White Ovals (WOs) are more frequent in the South hemisphere and very homogeneous in terms of their color which is nearly identical to the STrZ. The North hemisphere shows red and white ovals at the same latitude and an intermediate orange-white oval at high subpolar latitudes. The largest anticyclones, the GRS and oval BA, have a detailed color structure in their interior. Arrows in the scatter plot show the color-altitude structure of both ovals from their inner center to their outer rims. The reddest part of the GRS is located in the southern half of its inner core and the GRS becomes progressively less saturated in color at increasing distances from its center. This is a common behavior in color studies of the GRS (Simon-Miller et al. 2006; 2014; 2015). The center of oval BA is white and covered by high hazes that go down at the outer structure of the oval (Pérez-Hoyos et al. 2009). The oval is surrounded by a ring of faint red material and circled by more white material with even less upper haze content. The detailed color-altitude structure of the GRS and oval BA is further explored in higher spatial resolution in Figure 9 where the scatter diagram focuses on the radial structure of the vortices. The contrast with the environment and the radial distribution of colors are evident in both cases.

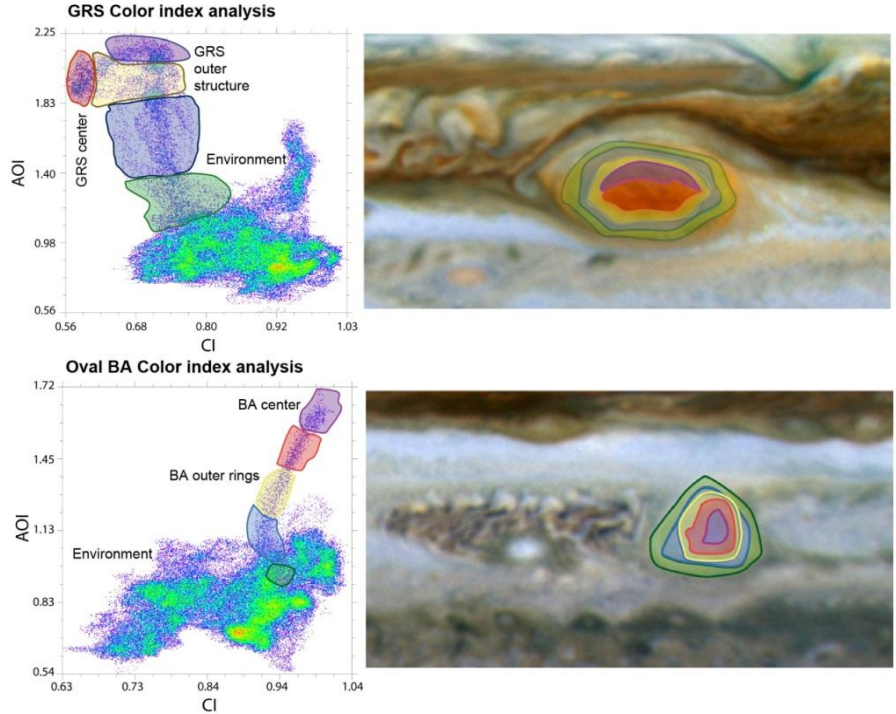

**Figure 9:** Scatter plot of color and altitude opacity indices for the Great Red Spot (top) and oval BA (bottom). Each plot is restricted to the region shown in the right panels.

While the overall relative positions of different vortices and structures in this color index diagram coincide with data presented by Sánchez-Lavega et al. (2013) the extreme values and specific numbers characterizing particular vortices differ in both works. For instance the AOI of the GRS in Sánchez-Lavega et al. (2013) was 1.5 for data acquired in 2008 while it is close to 2.0 in this work for data acquired in 2000. This is probably mainly caused by differences in the UV filters which have an effective wavelength and band passes in terms of the filter FWHM of 264 and 38 nm for Cassini ISS data and 255 and 43 nm for the HST data used in that work (besides other smaller differences such as the cameras CCD responses in those wavelength ranges), although a small color change in the GRS cannot be truly disregarded (Simon et al. 2014). Both works are fairly coincident in the values of the CI indices for similar structures except for oval BA which experienced a reddening in 2006 (Pérez-Hoyos et al. 2009; Wong al. 2011).

Colors and altitudes in Jovian clouds experience a continuity of values inside a given range as shown in Figure 8, instead of showing distinct well defined color clusters as investigated by Thompson et al. (1990). The large vortices (GRS and oval BA) have a very rich color-altitude radial structure that clearly manifests in Figure 9. Comparing the behavior of red cyclones in Figure 8 with results from the previous section demonstrates that they have the same red index as the GRS but a lower value of the AOI. This result does not allow to solve the debate between two chromophores versus one chromophore covered by high or low clouds.

**5. Jupiter colors in a CIE chromaticity diagram**

Astronomical color images have enhanced colors different to what a human observer would see through an eyepiece in a telescope. Evaluation of true-colors of astronomical objects as they would be seen by a human observer has a value on historical grounds to compare with color descriptions on eyepiece observations. In the case of Jupiter a relevant question is how red are the red features on the planet such as the GRS. Descriptions of true colors of Solar system objects are available in the literature but generally simplifying each body to its main color (Young 1984, 1985) or by simply comparing the reflectivity ratios in two broadband colors, a technique that has also been used to characterize the mean color of some exoplanets (Evans et al. 2013). Our goal here is to describe the variety of "true" colors in Jupiter clouds. Additionally, characterizing colors with a different technique might be useful to deepen or disregard the conclusions of previous analysis above. We will assume that a broadband red, green, blue filter photometric image would appear as a close match to the visual observer. This is only an approximation since: (1) Imaging filters have not the same response as the human eye. (2) Our red filter is a synthetic one that is an approximation to a more standard red filter. However the approximation is close enough to describe Jupiter range of colors and specially the color differences that can be found between different regions.

*5.1. Chromaticity*

Wyscezki and Stiles (1982) describe the field of colorimetry in detail. Human perception on color depends on three types of light sensitive cone cells with spectral responses that largely overlap (Bowmaker and Dartnall, 1980). Pure light of a single monochromatic wavelength stimulates differently the three types of cells. For this reason images formed by a trichromatic additive color space of three primaries (e.g. any RGB color space or image), pose difficulties for the definition of colors. One difficulty is that pure spectral colors imply negative values for at least one of the three primaries and can not be accurately represented in additive color spaces. To avoid negative RGB values,

"imaginary" primary colors and corresponding color-matching functions were formulated defining the widely used CIE 1931 color space. These imaginary primaries are denoted as X, Y, and Z. They are referred as tristumulus values and are used as a basis for defining colors precisely. The XYZ values are similar to the cone responses of the human eye and are defined so that Y takes the role of perceived luminance, Z is quasi-equal to blue stimulation and X is a linear combination of cone response curves chosen to be nonnegative (Wyscezki and Stiles, 1982). The three primaries are defined so that any color can be determined from using only positive values of the primaries. For a purely additive RGB nearly true color space, the X, Y, Z coordinates can be found from the following transformation (Wyscezki and Stiles, 1982; Fairman et al. 1997):

$$X = 0.49R + 0.31G + 0.20B$$
$$Y = 0.17697R + 0.81240G + 0.01063B \quad (11)$$
$$Z = 0.00R + 0.01G + 0.99B$$

Trichromatic x, y, and z coefficients are defined from them as

$$x = \frac{X}{X+Y+Z}$$
$$y = \frac{Y}{X+Y+Z} \quad (12)$$
$$z = \frac{X}{X+Y+Z}$$

These coefficients have the advantage that any color can be defined by the values of only $x$ and $y$ since $z$ is defined by the identity $x + y + z = 1$. Therefore, positive values of $x$ and $y$ define any color disregarding the amount of light/brightness.

*5.2. Chromaticity of Jupiter true colors*

We transformed the RED*, GRN and BL1 into tristumulus and chromaticity coordinates following eqs. (11) and (12) above. Figure 10 shows a scatter diagram of the chromaticity of our nearly true color map of Jupiter. The figure also displays the typical range of colors perceived by the human eye (the color gamut) and a map of the planet where particular features appearing in the color gamut of Jupiter colors are identified. Jovian colors stand close to the standard white (the so-called equal energy point E) showing the planet as a far less colored object when compared to published media depicting the planet in vivid colors (Young, 1984, 1985). Only faint hues of orange-yellow are found in this analysis in agreement with previous rough descriptions of true colors (Peek, 1958; Owen and Terrile, 1981). Most of the color variation is found across a single straight line with one of its extremes close to the standard white implying that most of Jupiter colors can be obtained by mixing white and a slightly orange material (therefore white ice particles and different quantities of a faint orange chromophore whose color is given by $x$=0.375 and $y$=0.370). However, there is a slight color variation in the perpendicular direction between these two points that requires an additional color to explain the gamut. This agrees with our previous conclusion regarding the need to incorporate a second chromophore at least in particular locations of the planet. In particular, this second chromophore is required to explain the color of the red cyclones in agreement with our previous results from PCA. A new result from this analysis is that this second chromophore should have a similar color to the human eye (although it would still be distinguishable as a different shade of red). Convective storms appear either white (in the SEB) or as a mixture of white material with some orange chromophore material typical of the environment in the NEB.

MacAdam ellipses are traditionally used to compare the precision of color matching for a human observer (MacAdam, 1942; Silberstein and MacAdam, 1945) and to define how well the human eye can distinguish color differences. Roughly speaking, an observer cannot distinguish differences in color from colors inside a given MacAdam ellipse in a chromaticity diagram. The size of the ellipses varies in different parts of the diagram and in the particular region covered by Jovian colors the ellipses have sizes of about 0.004 in $x$-$y$ coordinates allowing to distinguish the subtle color differences between the red hues of the belts and the GRS. MacAdam ellipses are shown in Figure 10 for the whole color gamut (ellipses are 10 times their real size following their traditional representation) and in the region covered by Jupiter colors (ellipses here are plot at their real size). In spite of Jupiter having faintly colored clouds, human perception of color is enough to distinguish the subtle color variations in the planet including the rich colored structure of large-scale features such as the GRS.

Finally, in terms of visual colors the whitest regions in the planet are indeed the SEB convective storms, the STrZ and WOs including most of oval BA. All of them reside very close to the Equal Energy point "E" ($x$=1/3, $y$=1/3) that can be considered as standard white. The reddest regions are the red cyclones while other red features such as the Great Red Spot and the North Equatorial Belt are slightly more orange.

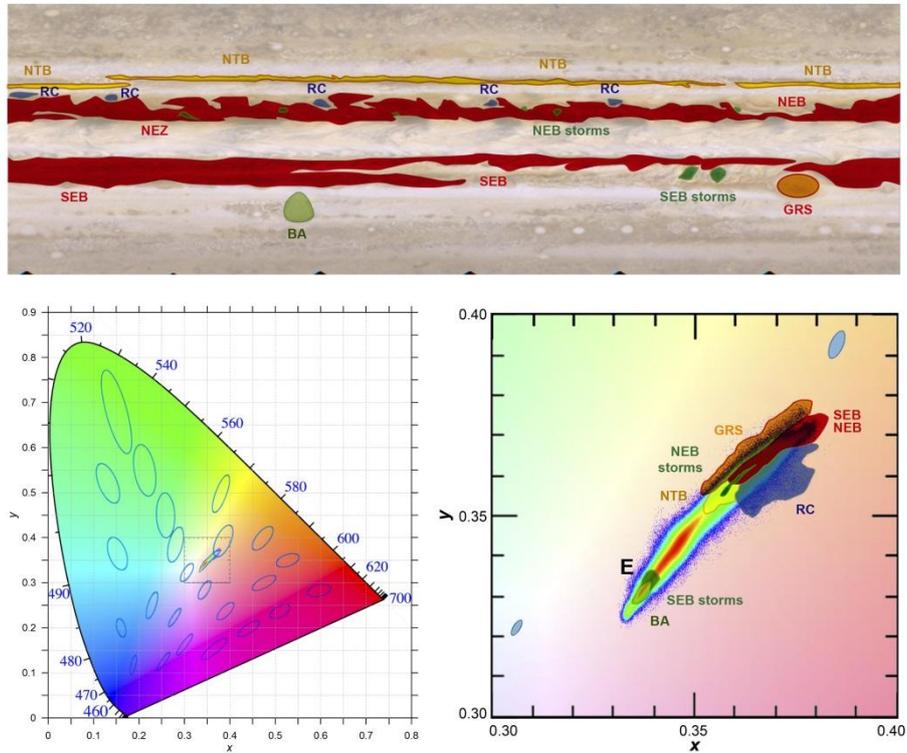

**Figure 10: Chromaticity of Jupiter colors.** Upper panel: Nearly true color map of Jupiter with colored regions highlighted. Bottom left panel: Full gamut of colors visible to the human eye in terms of the 1931 CIE chromaticity diagram. Purely spectra colors locate in the outer line of the diagram and are indicated for certain wavelengths. Jupiter colors locate in the x-y region bounded by the 0.3-0.4 values in both axes. Bottom right panel: Zoom into the region covered by Jupiter colors showing the scarce variation of color from different regions of Jupiter atmosphere as identified in the upper panel. The E point represents the "Equal energy point" (x=1/3, y=1/3). Standard MacAdam ellipses (1942) are plotted on both panels. Ellipses represent regions of colors indistinguishable to the human eye and measured from experiments with human observers. Ellipses are 10 times their actual length on the left bottom panel (following the conventional tradition in their representation) and their real length in the right bottom panel.

## 6. Conclusions

We explored the spatial distribution of colors, clouds and hazes in Jupiter following different techniques already tested in the literature: Principal Component Analysis, color indices and a chromaticity map. Additionally we computed a simplified map of methane optical depth at 890 nm that could help to understand part of the results. In all cases we examined cylindrical projections of the planet from observations obtained from the Cassini ISS during the early phase of the Jupiter flyby. Because of the viewing geometry our study is restricted to planetocentric latitudes ±55º.

- PCA strongly suggests the need to incorporate two chromophores or coloring processes to explain the diversity of visual colors in the planet. This agrees with conclusions from previous studies based on HST observations. The spatial distribution of a single main color agent explains the vast majority of shades of colors in the planet but a few small regions require a second coloring agent. The small brown cyclonic ovals (similar to the barges observed by the Voyagers) are anomalous regions in terms of their color. Also the convective regions in the NEB and the turbulent wake of the GRS show an anomalous behavior with very white particles that are interpreted as fresh material. Because of their spectral differences identified from PCA, these regions merit special attention in studies of colors in the planet aimed to identify the chemical nature of the red chromophore.
- We tried to use PCA to separate spatial information coming from the clouds and hazes. This proved not to be efficient because of the correlation of the visual aspect of the planet in images acquired in methane band filters and in images sensitive to the top of the ammonia cloud. A map of methane optical depth at 890 nm has been presented and compared with results from PCA. The map correlates well with PC 5 of the full PCA for the 9 images but it is not clear how the results from PCA could be used to infer other cloud properties or to separate the lower cloud and upper hazes.
- We investigated the colors of Jupiter clouds and altitudes of their hazes through color altitude indices based on image ratios. A full color-altitude diagram of the planet shows that some color-altitude clusters exist in Jupiter but that the colors and altitudes vary continuously from one region to other. The only well identifiable feature in the color-altitude diagram is the GRS, that stands apart of the rest of the planet. The structure of color and haze altitude for the GRS and oval BA is examined in detail and has a radial component.
- We characterize Jupiter true colors from a synthetic red image designed to match a wideband red observation and green and blue images. Subtle color differences between the belts and other red regions of the planet can be discerned without difficulties. In particular, the red color of the GRS and the different red hue of the red cyclones can be well distinguished from the rest of the planet by a visual observer. The fact that these two red colors lie in a perpendicular direction in the diagram to the first one suggests the need of incorporating a second chromophore to explain the visible colors in the image and their location in the red cyclones. Both shades of red are similar and both chromophores would have only subtle differences in color.

- None of these analyses suggests the need to incorporate more than 2 coloring processes to explain the range of colors observed in Jupiter clouds.

Identifying the chromophores that tint Jupiter clouds and how they relate to cloud altitudes may only be attainable by detailed high-spectral resolution spectra of different regions of the planet. These spectra should be obtained over regions small enough to distinguish the color variations that appear at spatial scales of 1,000-2,000 km characteristics of the convective storms and small cyclones that appear spectroscopically different in PCA. Ground-based observations aimed to that objective should have to resolve sources of 0.3'', requiring adaptive optics. The new Multi Unit Spectroscopic Explorer (MUSE) spectrograph on the VLT (Bacon et al. 2010) is well suited for that task. Accurate spatially resolved high-resolution spectra of the white regions (STrZ), red features (GRS and Belts) and the dark red cyclones are required to identify the sources of colors in the planet.

A similar analysis of Cassini ISS images of Saturn is possible. While color information is less apparent in Saturn's atmosphere, the upper hazes and lower clouds might be better separated in the images. The goldish-yellow colors of the South hemisphere and the blue colors in the northern hemisphere at the time of the Cassini arrival to the Saturn system in 2004 have been reversing over the years as the seasons have been inverting. The question of how many chromophores are required to explain these faint colors and how they relate to the seasons is a very intriguing one.

**Acknowledgments**

We gratefully acknowledge the work of the Cassini ISS team. This work was supported by the Spanish project AYA2012-36666 with FEDER support, Grupos Gobierno Vasco IT-765-13 and by Universidad del País Vasco UPV/EHU through program UFI11/55.

# Appendix

## Table A1: Selected images (November 14 2000)

| Image | Time (UT) | Filter | Sub-spacecraft longitude (System III) | Resolution (km/pix) |
|---|---|---|---|---|
| N1352887104 | 09:47:12 | UV1 | 325.88 | 526.00 |
| N1352893104 | 11:27:12 | UV1 | 25.72 | 525.25 |
| N1352899104 | 13:07:12 | UV1 | 85.89 | 524.51 |
| N1352905104 | 14:47:12 | UV1 | 145.13 | 523.76 |
| N1352911104 | 16:27:12 | UV1 | 205.34 | 523.01 |
| N1352917104 | 18:07:12 | UV1 | 265.25 | 522.26 |
| N1352887288 | 09:50:18 | BL1 | 327.16 | 262.99 |
| N1352893288 | 11:30:18 | BL1 | 276.20 | 262.61 |
| N1352899288 | 13:10:18 | BL1 | 87.91 | 262.24 |
| N1352905288 | 14:50:18 | BL1 | 147.02 | 261.87 |
| N1352911288 | 16:30:18 | BL1 | 207.07 | 261.49 |
| N1352917288 | 18:10:18 | BL1 | 267.67 | 261.12 |
| N1352887325 | 09:50:55 | GRN | 328.05 | 262.98 |
| N1352893325 | 11:30:55 | GRN | 279.14 | 262.61 |
| N1352899325 | 13:10:55 | GRN | 88.24 | 262.24 |
| N1352905325 | 14:50:55 | GRN | 147.17 | 261.87 |
| N1352911325 | 16:30:55 | GRN | 207.33 | 261.49 |
| N1352917325 | 18:10:55 | GRN | 267.59 | 261.12 |
| N1352887358 | 09:51:28 | MT1 | 327.53 | 262.98 |
| N1352893358 | 11:31:28 | MT1 | 268.24 | 262.61 |
| N1352899358 | 13:11:28 | MT1 | 207.91 | 262.23 |
| N1352905358 | 14:51:27 | MT1 | 147.94 | 261.86 |
| N1352911358 | 16:31:27 | MT1 | 88.72 | 261.49 |
| N1352917358 | 18:11:27 | MT1 | 284.91 | 261.11 |
| N1352887392 | 09:52:02 | CB1 | 327.99 | 262.98 |
| N1352893392 | 11:32:02 | CB1 | 28.49 | 262.61 |
| N1352899392 | 13:12:02 | CB1 | 88.72 | 262.23 |
| N1352905392 | 14:52:02 | CB1 | 147.94 | 261.86 |
| N1352911392 | 16:32:02 | CB1 | 207.91 | 261.49 |
| N1352917392 | 18:12:02 | CB1 | 268.24 | 261.11 |
| N1352887210 | 09:48:59 | MT2 | 326.26 | 262.99 |
| N1352893210 | 11:28:59 | MT2 | 277.37 | 262.62 |
| N1352899210 | 13:08:59 | MT2 | 86.93 | 262.24 |
| N1352905210 | 14:48:59 | MT2 | 146.05 | 261.87 |
| N1352911210 | 16:28:59 | MT2 | 206.20 | 261.50 |
| N1352917210 | 18:08:59 | MT2 | 266.46 | 261.12 |
| N1352887255 | 09:49:45 | CB2 | 326.49 | 262.99 |
| N1352893255 | 11:29:45 | CB2 | 270.23 | 262.61 |
| N1352899255 | 13:09:45 | CB2 | 88.03 | 262.24 |
| N1352905255 | 14:49:45 | CB2 | 146.16 | 261.87 |
| N1352911255 | 16:29:45 | CB2 | 206.39 | 261.49 |
| N1352917255 | 18:09:45 | CB2 | 266.56 | 261.12 |
| N1352887145 | 09:47:54 | MT3 | 325.79 | 526.00 |
| N1352893145 | 11:27:54 | MT3 | 265.51 | 525.25 |
| N1352899145 | 13:07:54 | MT3 | 85.52 | 524.50 |
| N1352905145 | 14:47:53 | MT3 | 145.94 | 523.75 |
| N1352911145 | 16:27:53 | MT3 | 205.21 | 523.01 |
| N1352917145 | 18:07:53 | MT3 | 265.84 | 522.26 |
| N1352887174 | 09:48:24 | CB3 | 325.36 | 262.99 |
| N1352893174 | 11:28:24 | CB3 | 259.07 | 262.62 |
| N1352899174 | 13:08:24 | CB3 | 86.01 | 262.25 |
| N1352905174 | 14:48:23 | CB3 | 145.16 | 261.87 |
| N1352911174 | 16:28:23 | CB3 | 205.31 | 261.50 |
| N1352917174 | 18:08:23 | CB3 | 265.50 | 261.13 |

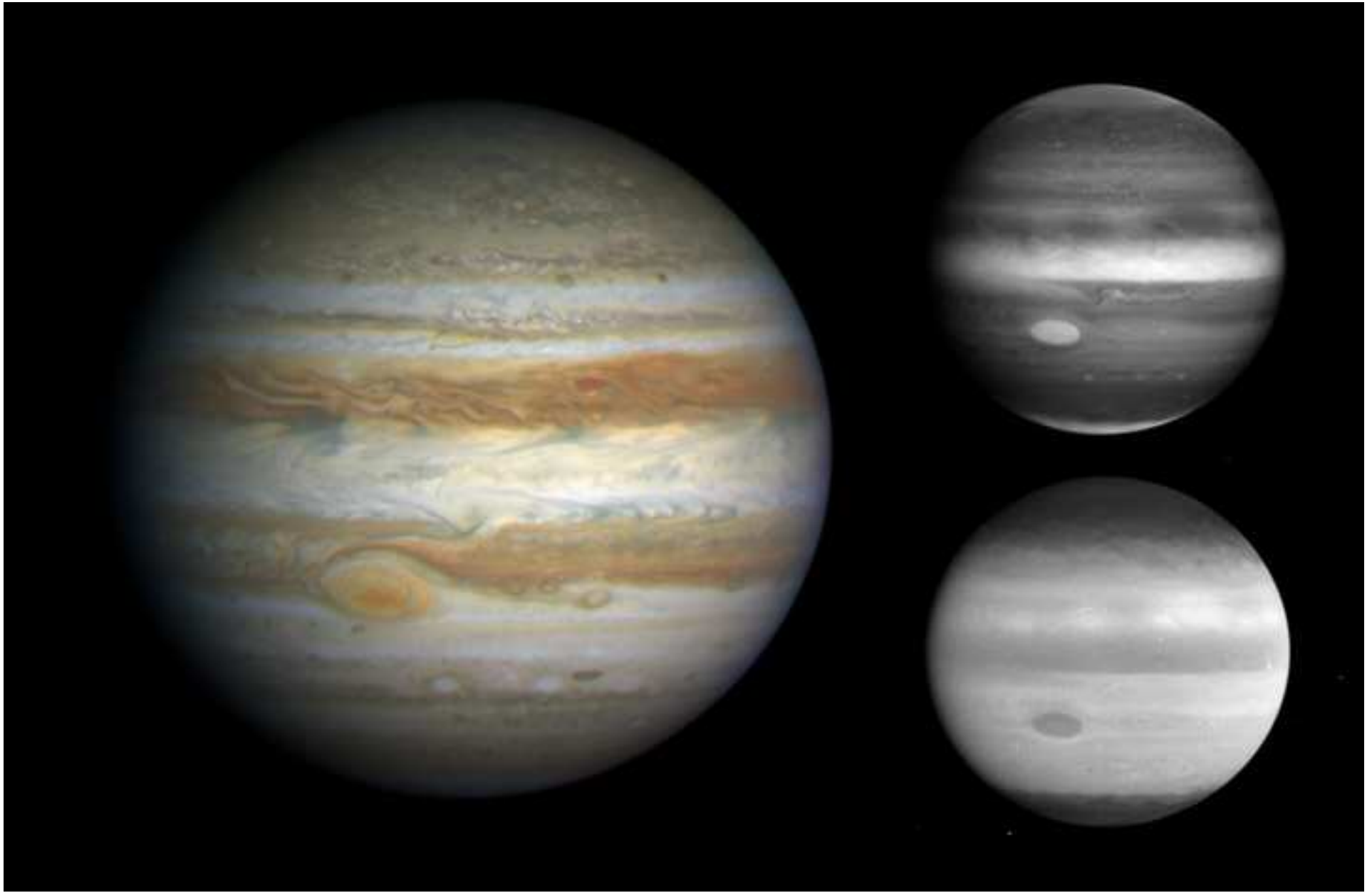

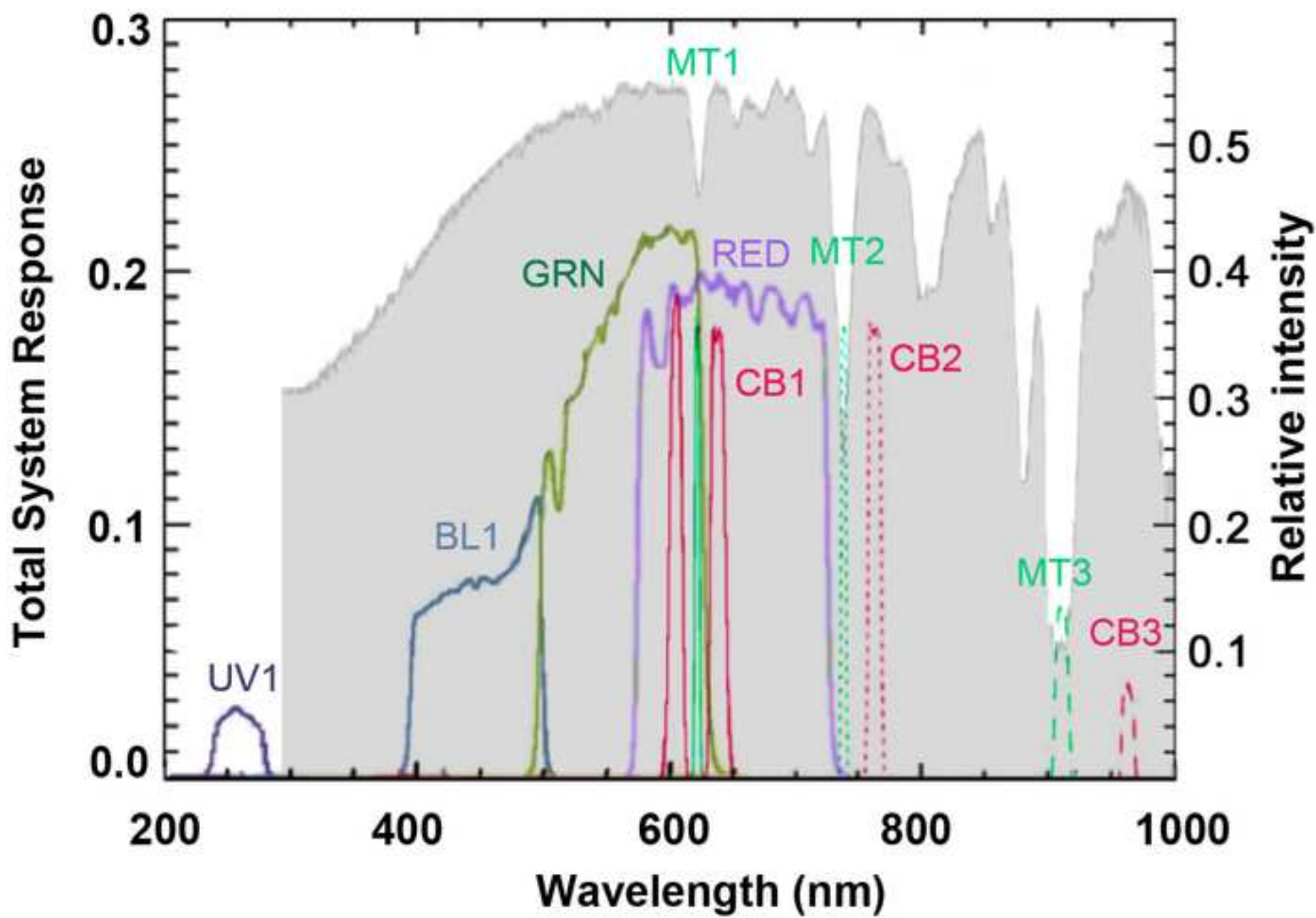

Figure03
Click here to download high resolution image

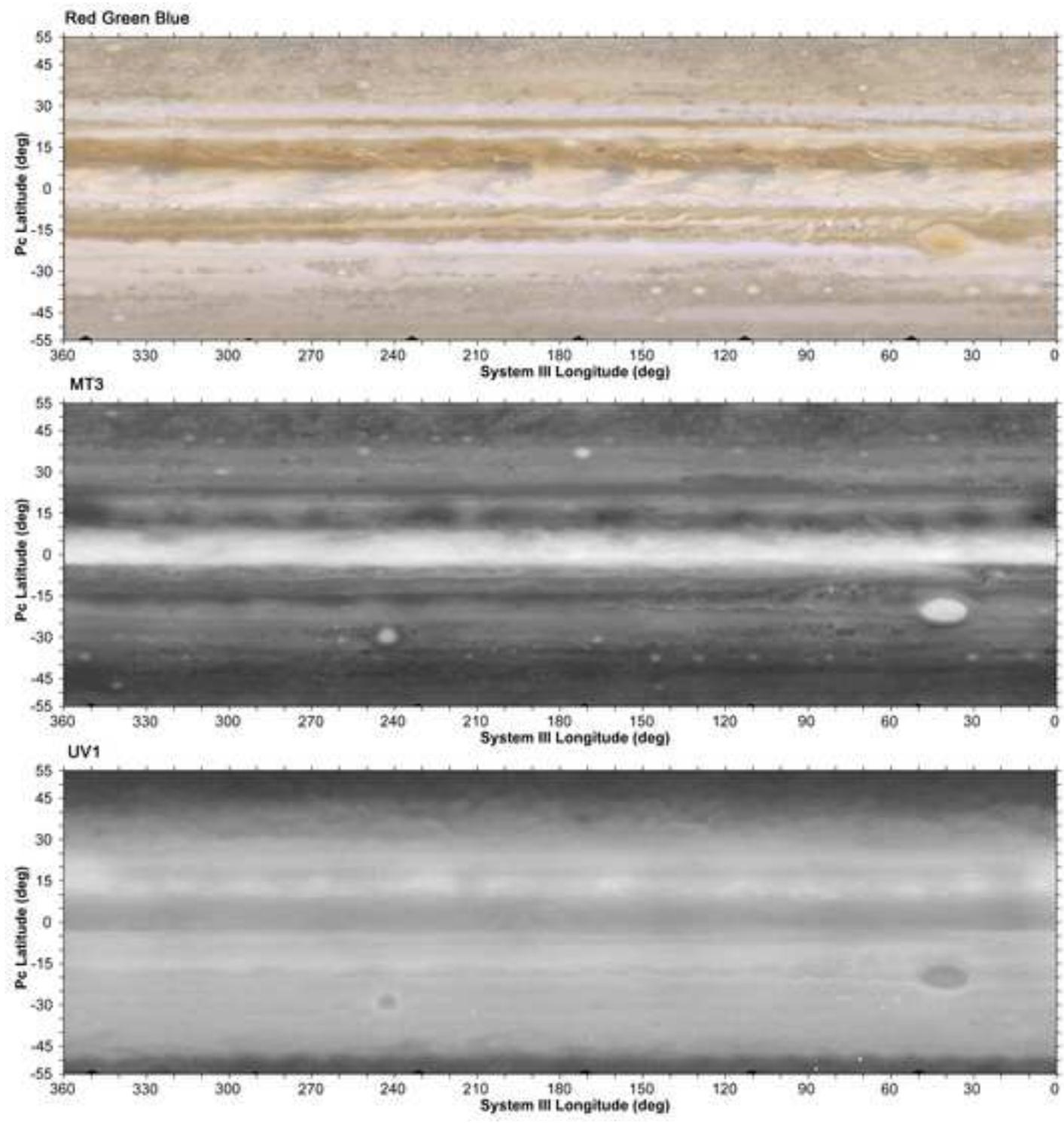

**Figure04A**
[Click here to download high resolution image](#)

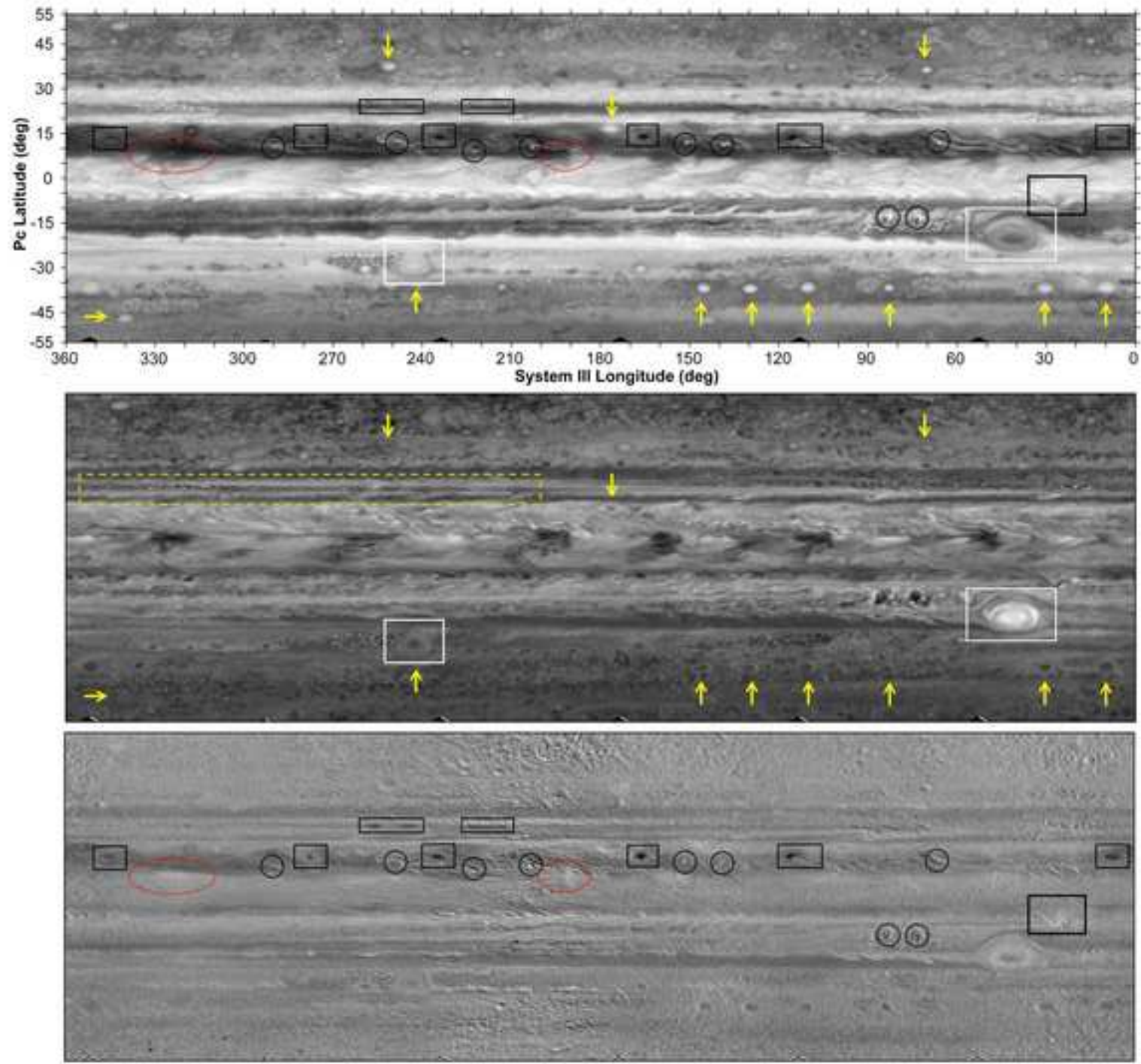

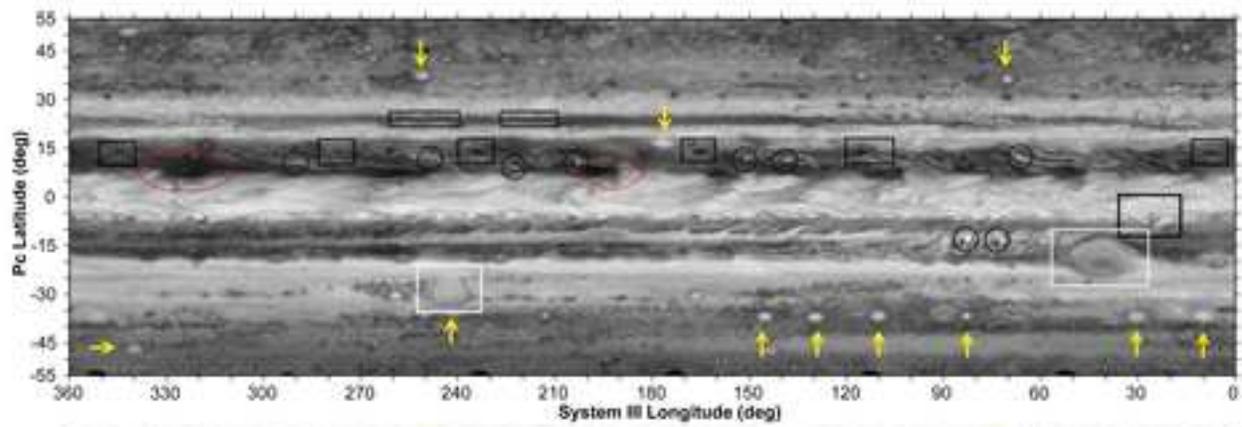
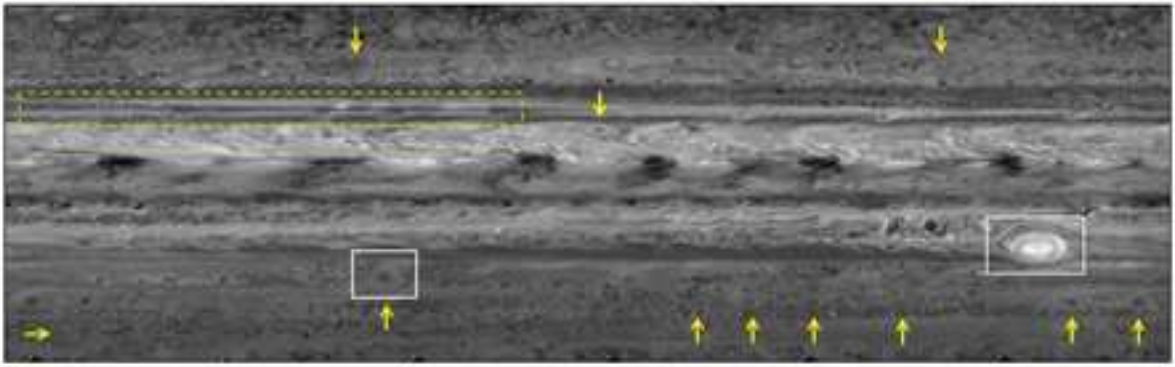
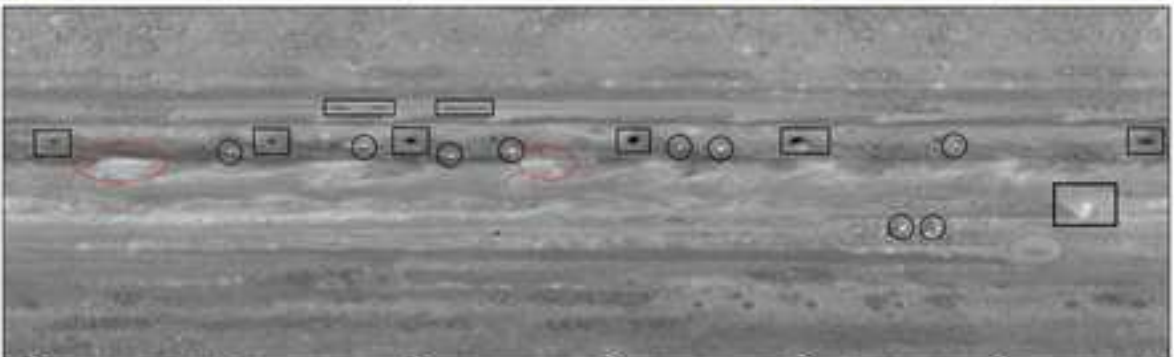
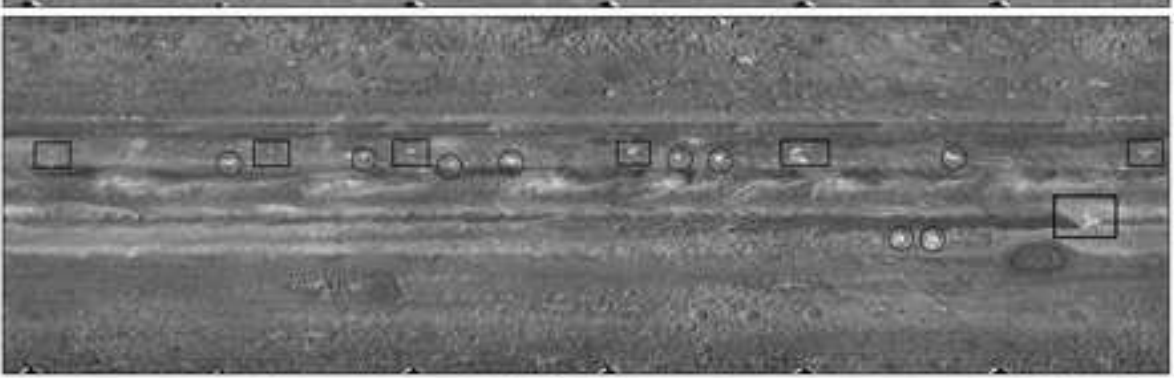
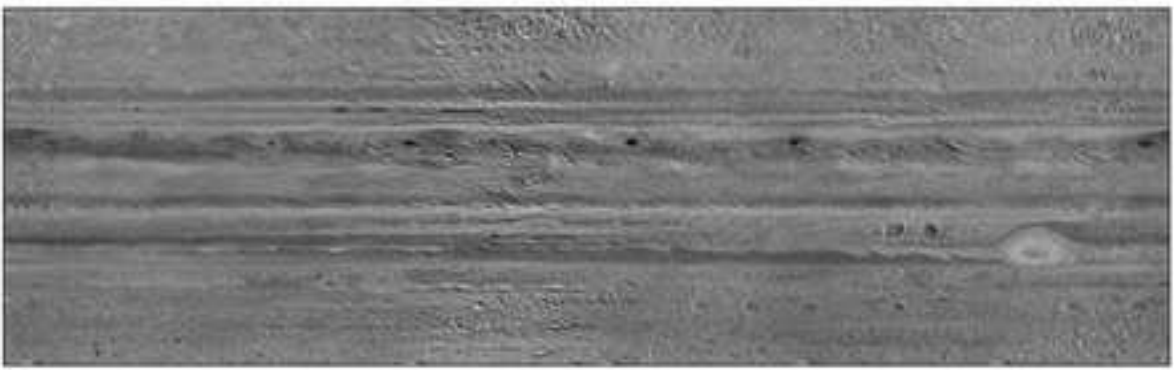

**Figure04C**
Click here to download high resolution image

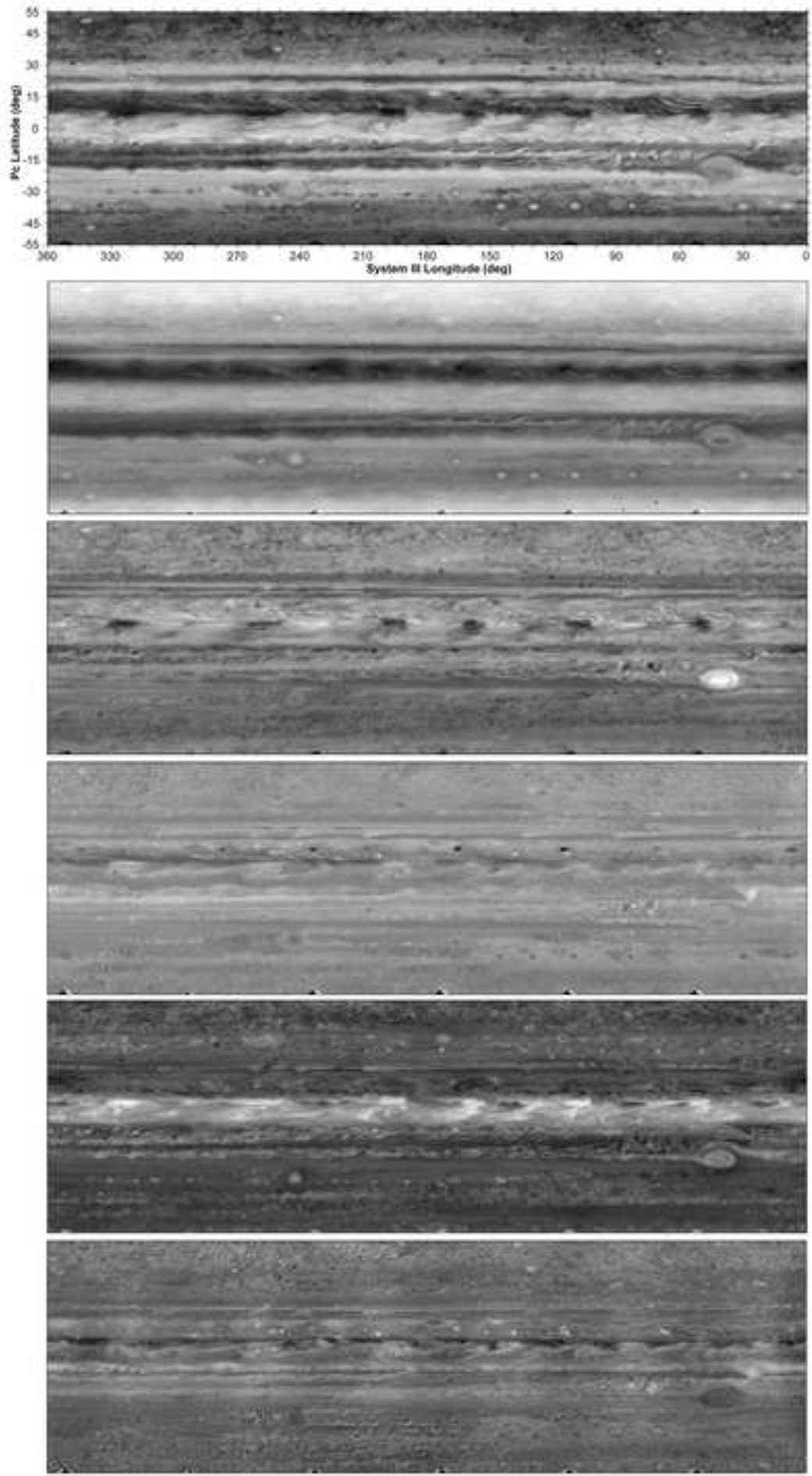

**Figure05**
[Click here to download high resolution image](#)

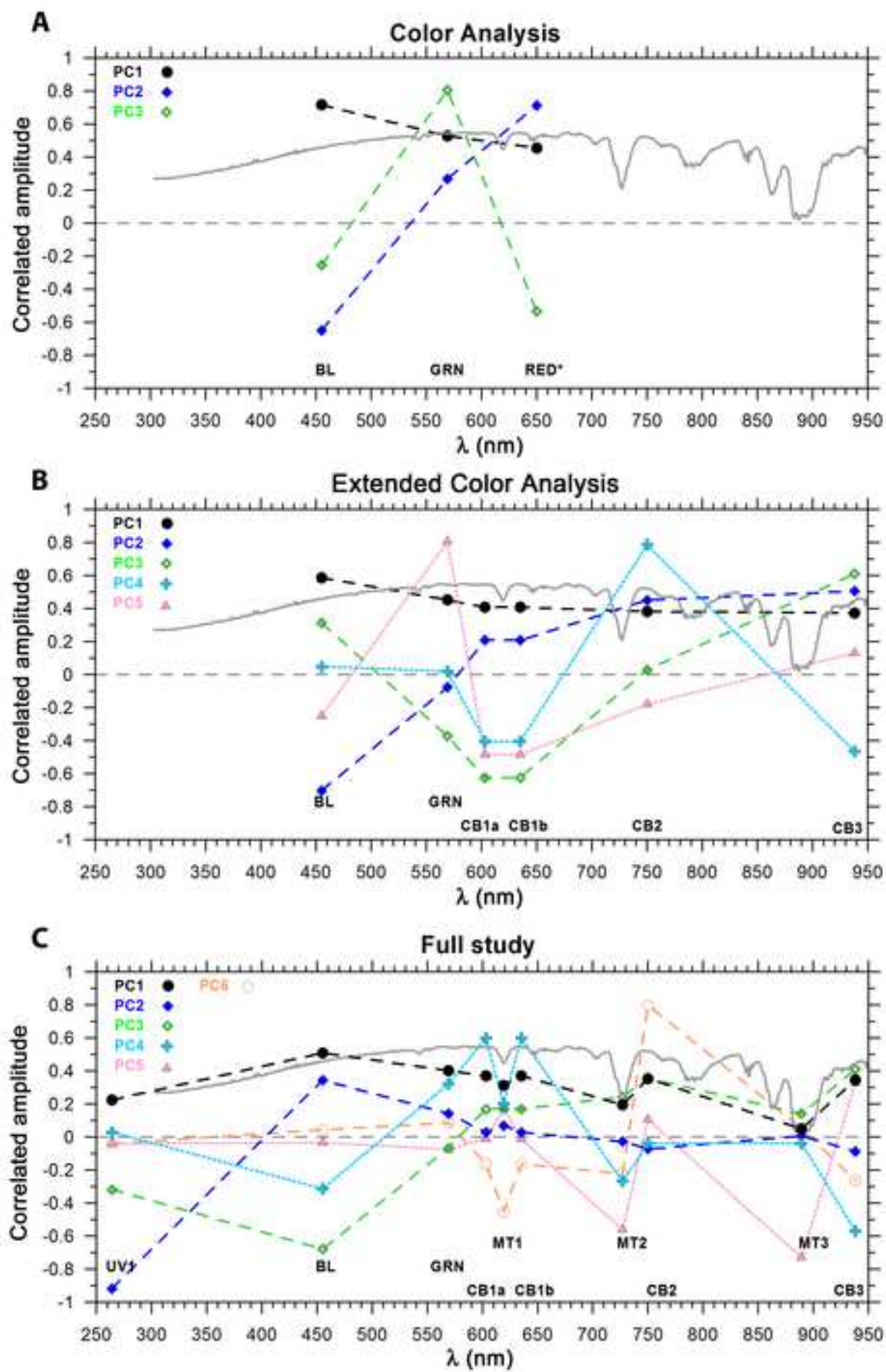

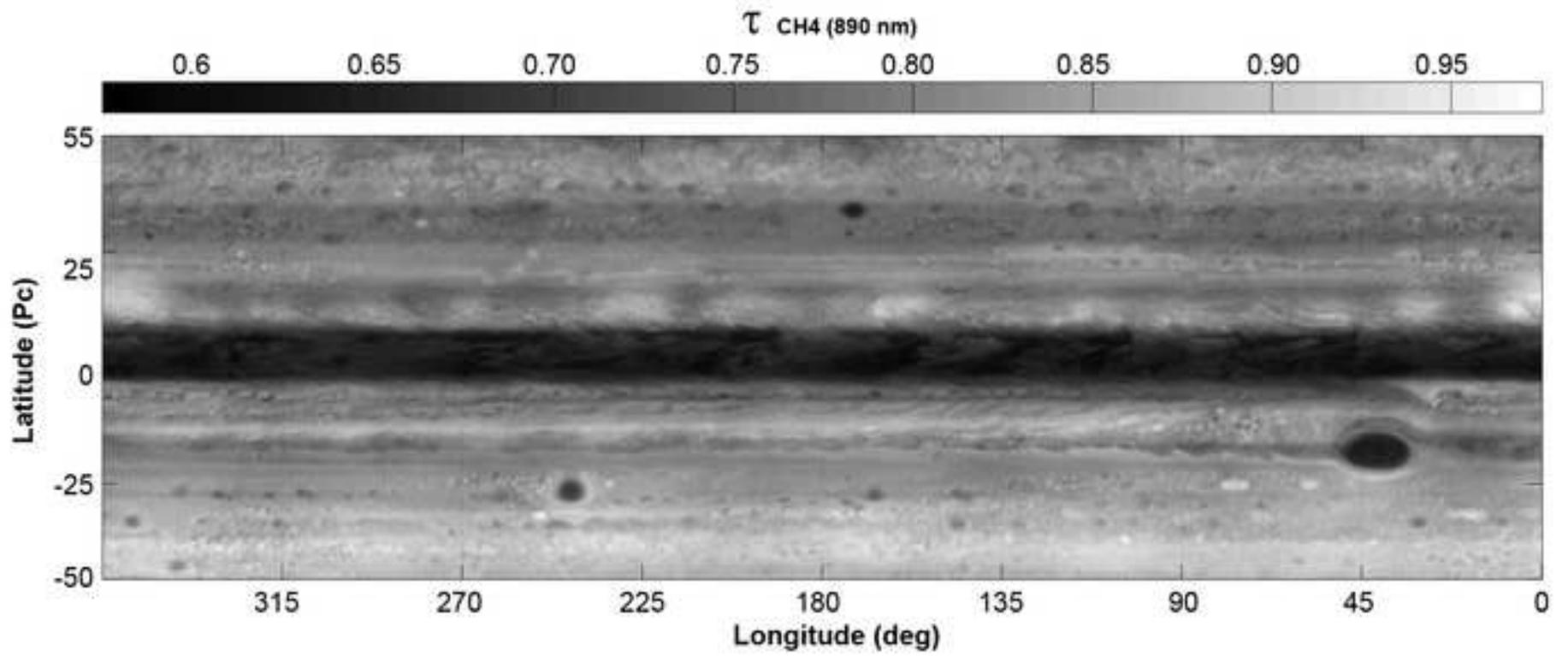

**Figure07**
**Click here to download high resolution image**

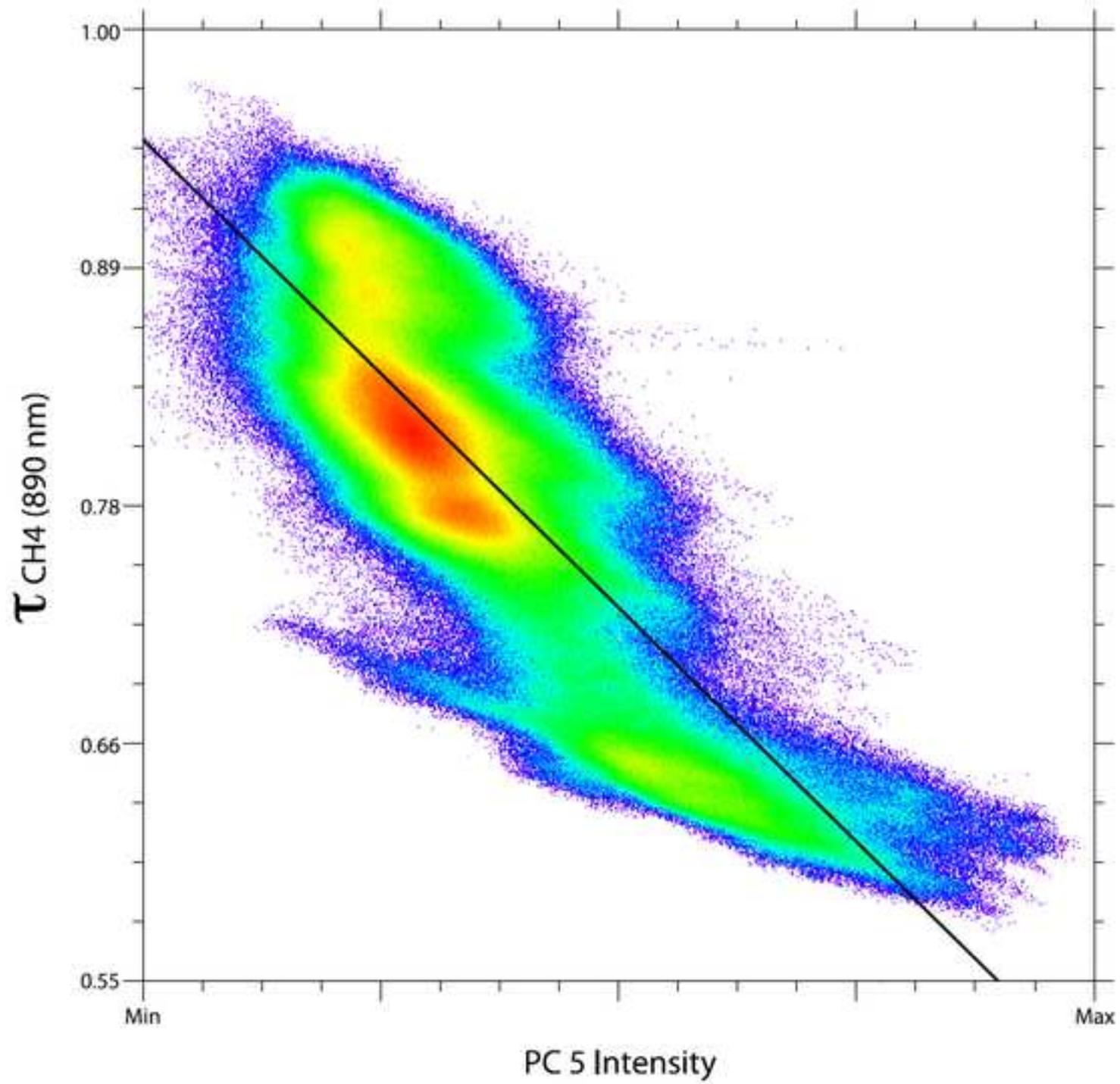

**Figure08**
**Click here to download high resolution image**

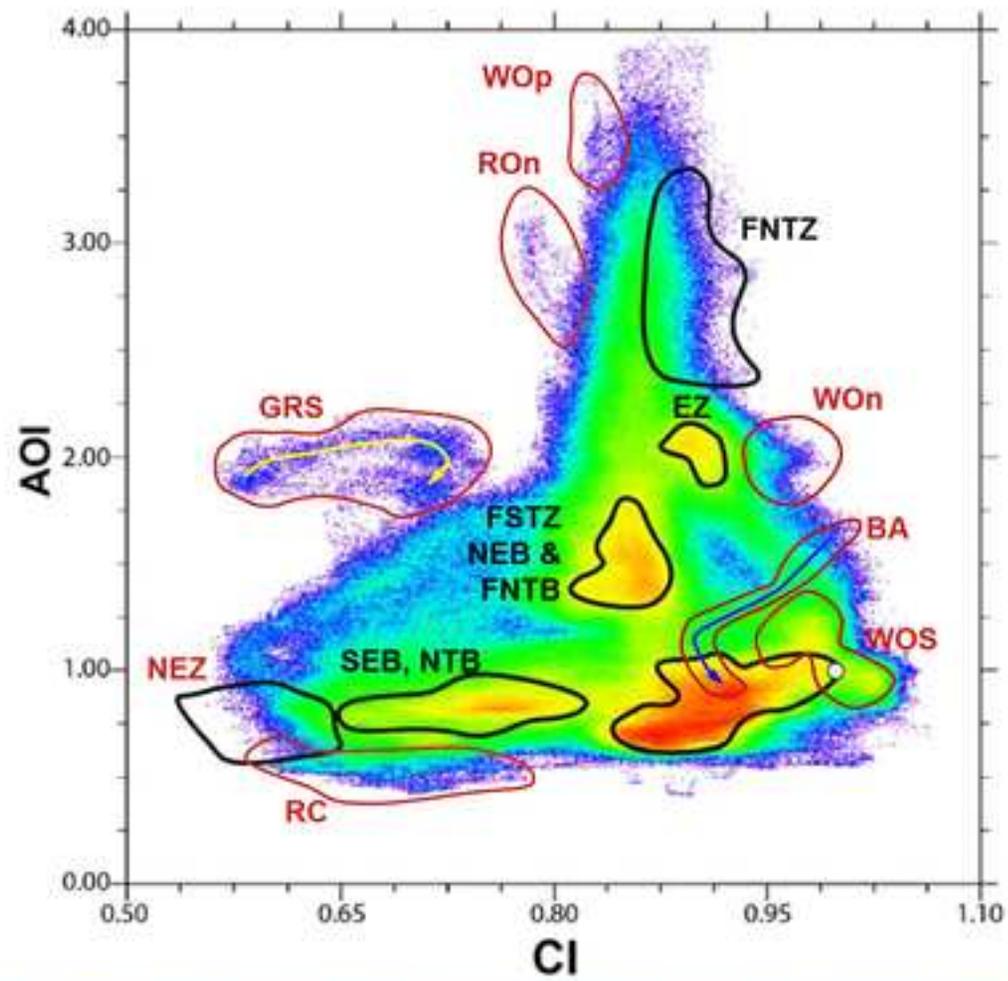
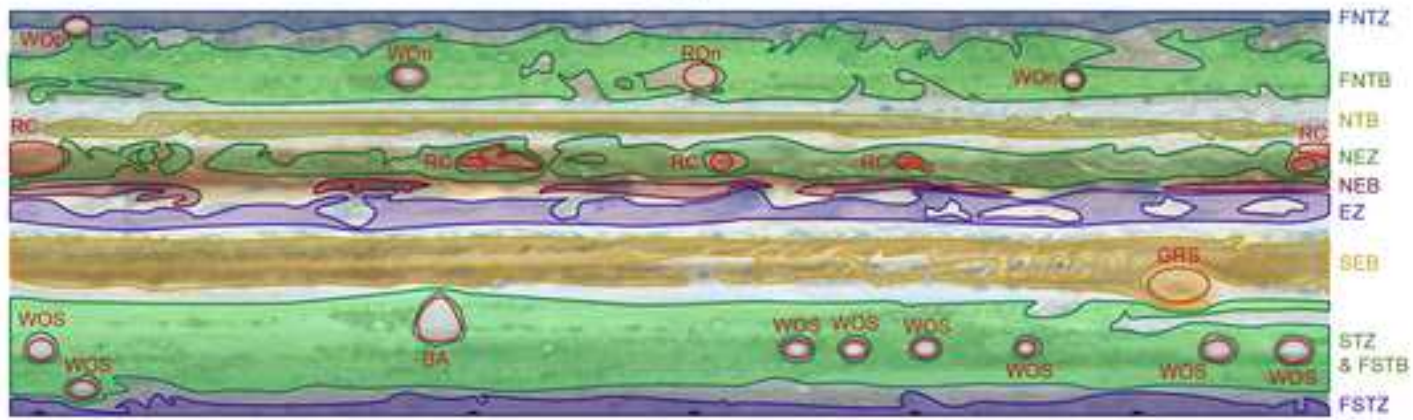

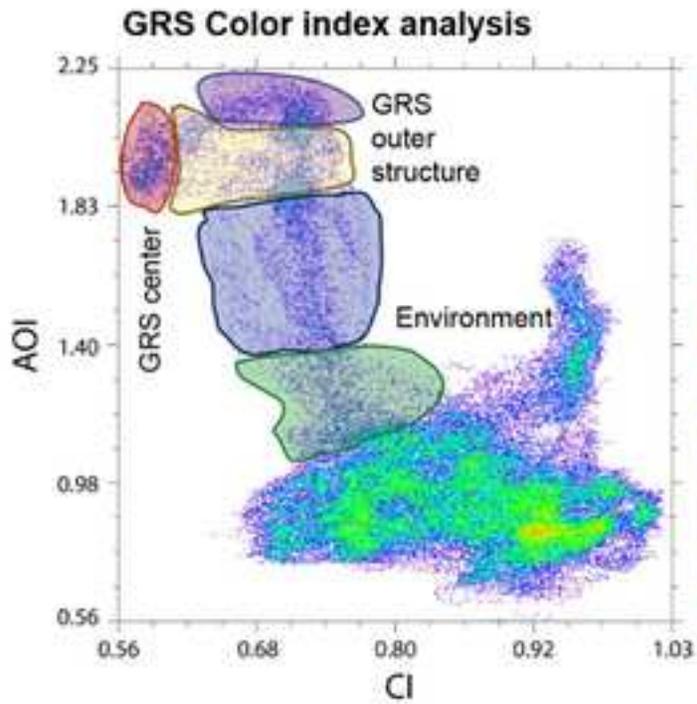
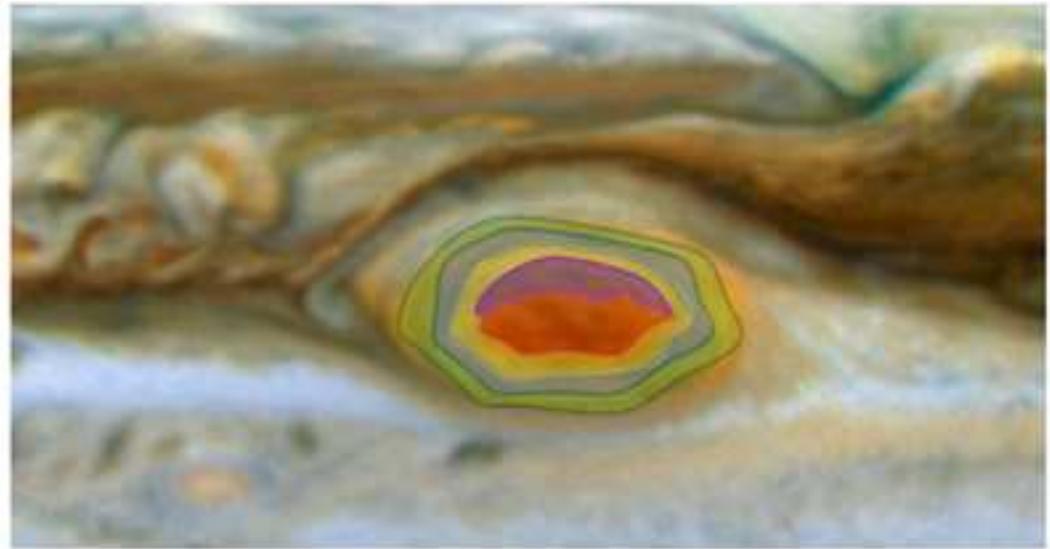
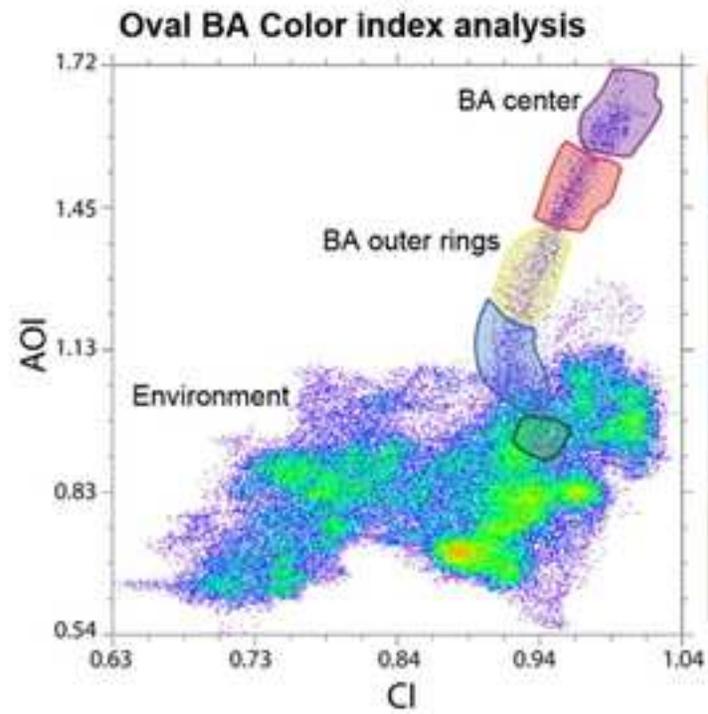
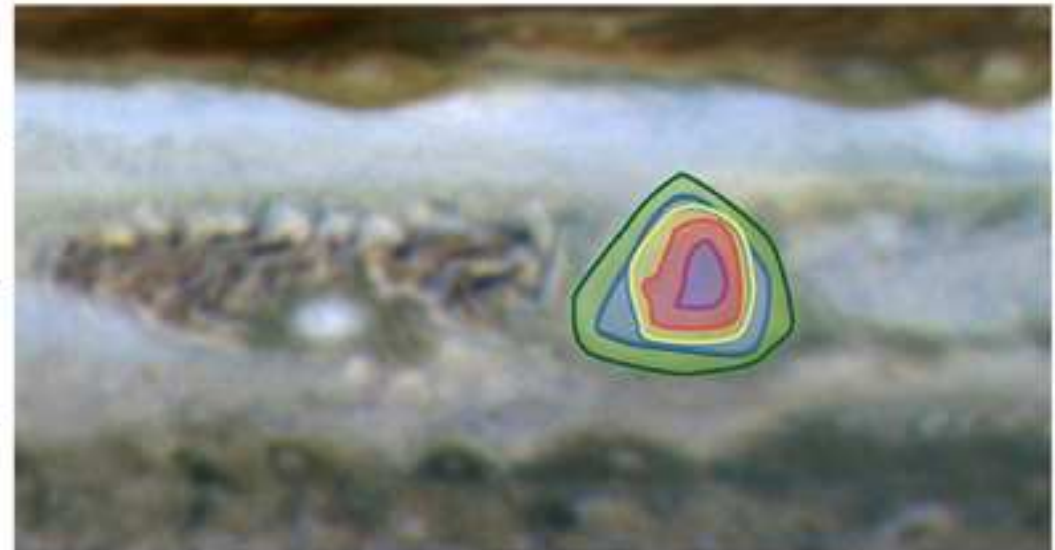

**Figure10**
[Click here to download high resolution image]

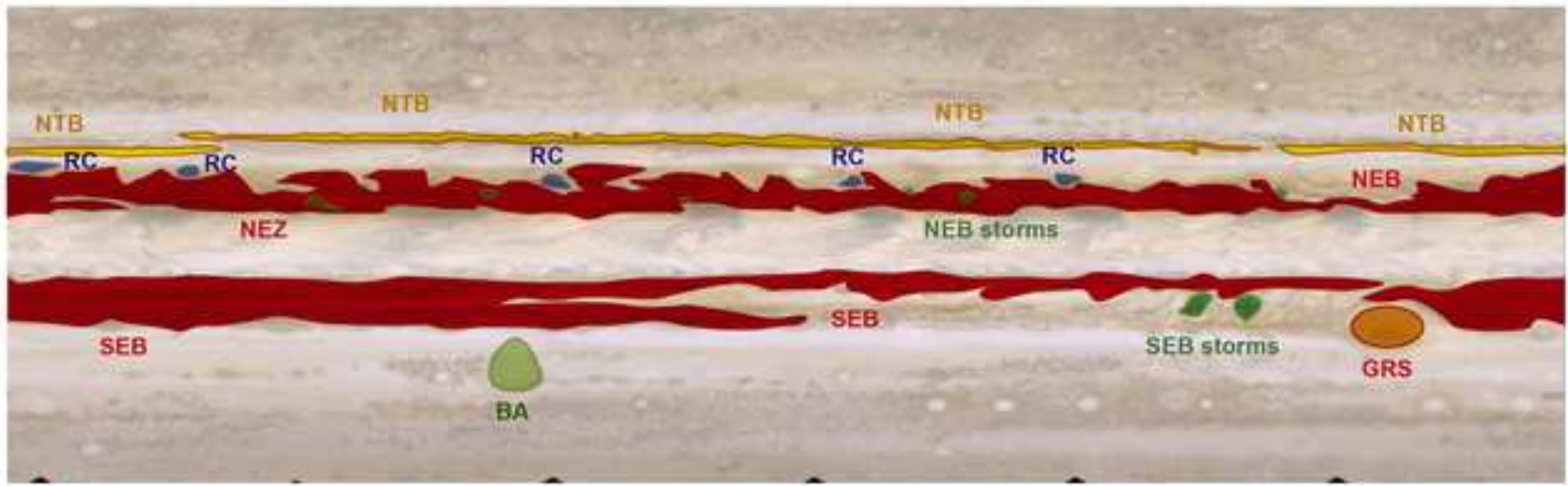
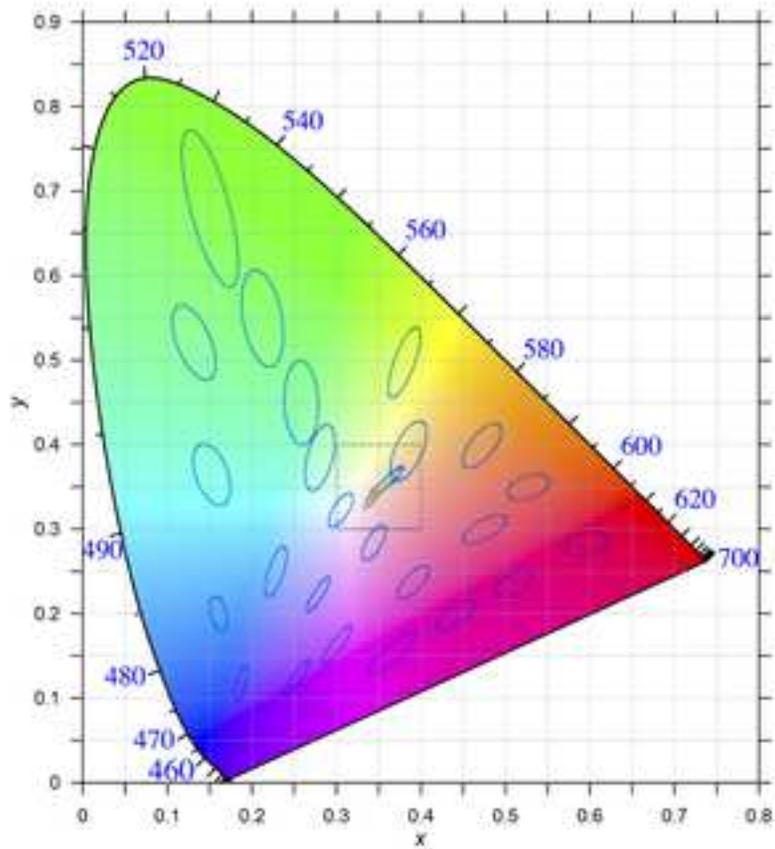
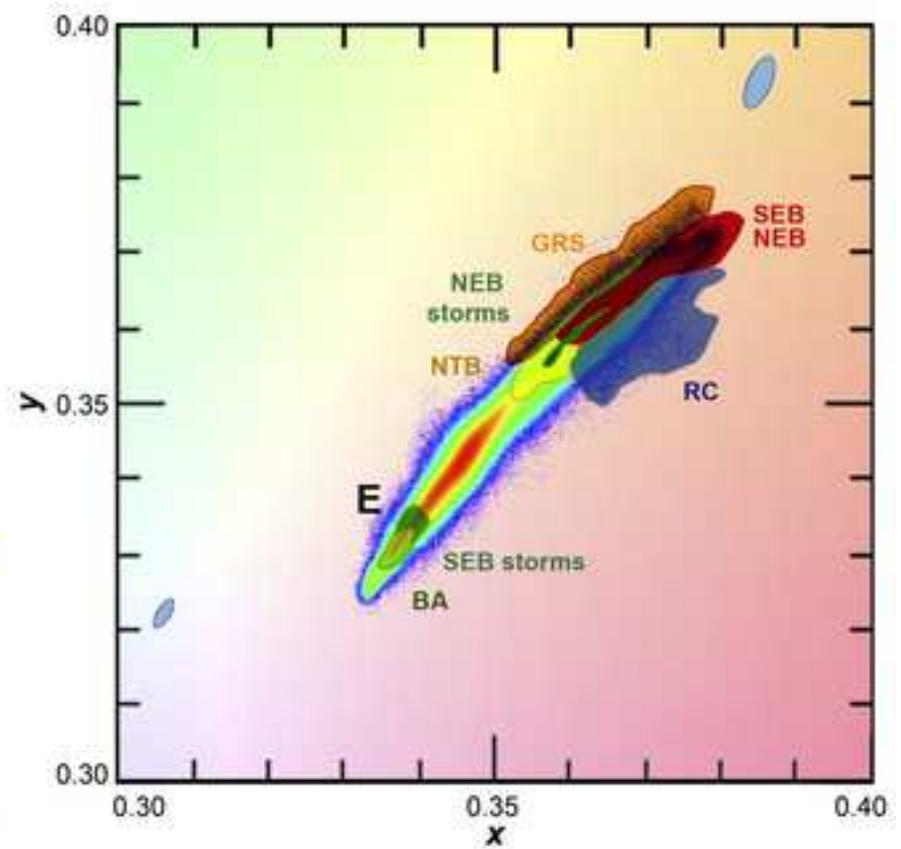